\documentclass{aa} 
\usepackage{graphicx}
\usepackage{txfonts}
\usepackage{natbib}
\usepackage{longtable}
\usepackage[figuresright]{rotating}
 \usepackage{caption}
 \usepackage{subcaption}
\usepackage{color}
\usepackage{subfig}
\usepackage{tabularx}
\bibpunct{(}{)}{;}{a}{}{,}
\usepackage{verbatim}

\newcommand{\Teff}{T_{\rm eff}}

\newcommand{\Dg}{\Delta_{\rm NLTE} \log g}
\newcommand{\Dm}{\Delta_{\rm NLTE} \rm{[Fe/H]}}
\newcommand{\nod}{---}
\begin{document}
\title{LTE or non-LTE, that is the question}
\subtitle{The NLTE chemical evolution of Sr in extremely metal-poor stars}
\author{C. J. Hansen\inst{1}, M. Bergemann\inst{2}, G. Cescutti\inst{3}, P.
Francois\inst{4}, A. Arcones\inst{5}, A. I. Karakas\inst{6}, K. Lind\inst{2},
and C. Chiappini\inst{3}}
\institute{Landessternwarte, ZAH, K\"onigstuhl 12, 69117 Heidelberg, Germany\\
\and 
Max-Planck Institute for Astrophysics, Karl-Schwarzschild Str. 1, 85741,
Garching, Germany \\
\and
Leibniz-Institut f\"ur Astrophysik Potsdam (AIP), An der Sternwarte 16, D-14482
Potsdam, Germany\\
\and
GEPI, Observatoire de Paris, CNRS Universite Paris Diderot, Place Jules Janssen,
92190 Meudon, France\\
 \and
TU Darmstadt, Institut f\"ur Kernphysik Theoriezentrum, Schlossgartenstr. 2,
64289 Darmstadt, Germany\\
\and
Research School of Astronomy \& Astrophysics, Mount Stromlo Observatory,
Weston Creek ACT 2611, Australia
}

\date{Received 18/10/2012 / Accepted 24/12/2012}
\abstract
{Strontium has proven itself to be one of the most important neutron-capture
elements in the study of metal-poor stars. 
Thanks to the strong absorption lines of Sr, they can
be detected even in the most metal-poor 
stars and also in low-resolution
spectra.
However, we still cannot explain the large star-to-star abundance
scatter we derive for metal-poor stars.}
{Here we compare Galactic chemical evolution (GCE) predictions with improved abundances for \ion{Sr}{i}
  and \ion{Sr}{ii}, including updated atomic data, to evaluate possible
explanations for the large star-to-star scatter at low metallicities.}
{We have derived abundances under both local thermodynamic equilibrium (LTE) and non-LTE (NLTE) for stars spanning a large interval of metallicities, as well as a broad range of other stellar
parameters. Gravities and metallicities are also determined in NLTE. We employed MARCS stellar
atmospheres and MOOG for the LTE spectrum synthesis, while MAFAGS and DETAIL
were used to derive the NLTE abundances. We verified the consistency of the two methods in LTE.}
{ We confirm that the ionisation equilibrium between \ion{Sr}{I} and
\ion{Sr}{II} is satisfied under NLTE but not LTE, where the difference between
neutral and ionised Sr is on average $\sim 0.3$\,dex. We show that the NLTE
corrections are of increasing importance as the metallicity
decreases. For the stars with [Fe/H] $ > -3$, the \ion{Sr}{i} NLTE correction is $\sim 0.35/0.55$\,dex in
dwarfs/giants, while the \ion{Sr}{ii} NLTE correction is
$<\pm0.05$\,dex.
}
{On the basis of the large NLTE corrections to \ion{Sr}{i}, \ion{Sr}{i} should
not be applied as a chemical tracer under LTE, while it is a good tracer under
NLTE. \ion{Sr}{ii}, on the other hand, is a good tracer under both LTE and NLTE
(down to [Fe/H] $\sim-3$), and LTE is a safe assumption for this majority species
(if the NLTE corrections are not available). However, the Sr abundance from
\ion{Sr}{ii} lines depends on determining an accurate surface gravity, which can be obtained from the NLTE spectroscopy of Fe lines or from parallax measurements. We could
not explain the star-to-star scatter (which remains under both LTE and NLTE) by the use of the Galactic chemical evolution
model, since Sr yields to date have been too uncertain to draw
firm conclusions. At least two nucleosynthetic production sites seem necessary to
account for this large scatter. }

\titlerunning{Sr abundances in late-type stars}
\authorrunning{C. J. Hansen et al.}
\maketitle
%
%
\section{Introduction}
Strontium is one of the two neutron-capture elements (namely strontium and
barium - Ba) that intrinsically show very strong absorption lines even in
metal-poor stars. 
At solar metallicity Sr is synthesised by a variety of nucleosynthetic sources
including the weak slow neutron-capture (s-) process that occurs in massive
stars \citep[e.g.][]{heil,pigna} and in AGB stars \citep{Trav}. In comparison, the
production of Ba is dominated by the s-process occurring in low-mass AGB stars
\citep{kaep89,busso,chrisrev}.
This picture changes at low metallicity, where Ba may be formed by a main rapid
neutron-capture process and Sr by a charged
particle process \citep{hoffman}. Additionally, very metal-poor rapidly rotating
massive stars might be significant producers of Sr and Ba via the s-process \citep{pignatari,chiapSr,frisch}. This may modify the presumption of pure r-process patterns in ultra metal-poor
(UMP; $-5 <$ [Fe/H] $<-4$ cf. \citealt{beersFe}) stars. For this reason, disentangling the nucleosynthetic origin of Sr and Ba in metal-poor stars would help us understand the formation and evolution of the
early Galaxy.

Only the 4077 \AA\, \ion{Sr}{II} line remains detectable in both
dwarfs and giants both in high- and low-resolution spectra of
metal-poor stars. Studying this line thus provides unique insight into
the behaviour of neutron-capture elements at all metallicities and
spectral resolutions, ranging from the low-resolution LAMOST survey to the
high-resolution Gaia-ESO survey\footnote{The \ion{Sr}{ii} line is detectable if a blue setting is used for Gaia-ESO follow-up observations.}. 
Clearly, accurate abundances are needed to fully comprehend
the chemical evolution of Sr. This means that the effects of non-local thermodynamic
equilibrium (NLTE) and deviations from hydrostatic equilibrium (3D) must be
taken into account in element abundance calculations, if we want to extract the
correct information from the future surveys' large flow of data. Clearly, such
calculations are a challenge, as was recently demonstrated for O \citep{gonher},
Ca \citep{spiteCa}, Fe \citep{bergSr}, and for Ba \citep{Ba3d}, while
\citet{bonifaciodw} provide 3D corrections for a large number of elements for
dwarfs. 
The estimates of 3D effects for the \ion{Sr}{ii} resonance lines have
been provided by \cite{collet}, who performed LTE calculations with 3D
radiative-hydrodynamics simulations of stellar convection for the metal-poor
stars. The 3D LTE corrections are of the order of $-0.15$ dex, with
respect to 1D LTE. The NLTE abundances of Sr were reported in a few studies, such as
those of \citet{srnlte}, \cite{mash99}, \cite{andrSr}, and \citet{bergSr}. For the
  \ion{Sr}{ii} resonance line, the NLTE abundance corrections are not large, typically within $\pm 0.2$ dex\footnote{For very
metal-poor (VMP; $-3 < $ [Fe/H] $<-2$) stars the corrections are less than about $\pm 0.08$ and
slightly greater for extremely metal-poor stars ([Fe/H] $<-3$).}, \textit{however,
they are sensitive to variations in the stellar parameters} \citep{bergSr}. Full 3D
NLTE calculations for Sr have not been performed yet.

Using the NLTE technique presented in \cite{bergSr}, we have now expanded the
stellar sample in order to study the chemical evolution of Sr in the Galaxy. We have
derived NLTE Sr abundances and NLTE stellar parameters for a sample of 21 stars,
plus comparison samples (51 very metal-poor stars from \citealt[][]{francois} and \citealt{bonifaciodw}). 
We also
include the predictions of the homogeneous chemical
evolution model for the Galactic halo of \citet{chiappini08},
computed with the most up-to-date Sr yields available in the
literature. Given the still large uncertainties on the stellar yields, the goal
of comparing the data and the chemical evolution model is to give a first
impression of how far the available stellar yields are from explaining
the data. A comparison with inhomogeneous chemical evolution
models, more suitable to low-metallicity environments, is beyond the scope of
the present paper, see \citet[][]{cescutti,cescut10}.
The paper is structured as follows. Sections 2 and 3 describe the observations,
stellar parameters, and NLTE calculations, respectively. Sections 4 and 5 present
the results, yields, and the chemical evolution model. Conclusions can be found
in Sect. 6.

\section{Sample and data reduction}

A sub-sample of stars (marked `u' in Sect. \ref{results}) was taken from
\citet{hansen}. These stars were observed with UVES/VLT \citep[][$R \geq
40000$]{dekker} between 2000 and 2002, and their spectra were reduced with
the UVES pipeline (v. 4.3.0). The spectra have a signal-to-noise ratio, S/N, $>
100$ at 3200\,\AA. Two stars (see `h' in the following) were observed
with HIRES/Keck \citep[$R\sim 60000$,][]{vogt94}. Their spectra are of similar
quality as the UVES spectra, and they have been retrieved from the HIRES reduced
data archive. The pipeline-reduced data were wavelength-shifted, co-added, and
had their continua normalised before our analysis. For further details we refer
to \citet{hansen}.

Three stars (\object{HD 134169}, \object{HD 148816}, \object{HD 184448}) were
observed with the FOCES echelle spectrograph at the 2.2\,m telescope of the CAHA
observatory on Calar Alto, during 1999 and 2000, and were kindly made available
to us by T. Gehren. The spectra have a resolution of $\sim 60000$ and an $S/N$
of $\sim 200$ near 5000 \AA. More details on the observations and data reduction
can be found in \citet{gehren04,gehren06}.

We selected the stars according to the following criteria: 1) the observations
cover
the spectral range of the \ion{Sr}{i} 4607\,\AA\, line, 2) accurate photometry
is
available, and 3) the stars cover a broad stellar parameter space to test the
\ion{Sr}{i} and \ion{Sr}{ii} abundance behaviour at different temperatures,
gravities, and metallicities. Our sample thus consists of 21 dwarf, sub-giant,
and giant stars. Here we have disregarded carbon enhanced
metal-poor stars (CEMP-s/CEMP-no --- with and without s-process overabundances).
Since the CEMP-s stars tend to have very large s-process abundances, we
are slightly biased against high Sr abundances at low metallicity. 

For comparison, we include extremely metal-poor stars from \cite{francois} and
\cite{bonifaciodw}. For details about the observed data, we refer to these
publications. 

\section{Methods}

In this work, NLTE effects are accounted for in the determination of basic
stellar parameters (surface gravity and metallicity), as well as Sr abundances.
We describe the analysis in detail below.

\subsection{Model atmospheres}
\begin{figure}[ht!]
\centering
\includegraphics[width=0.48\textwidth]{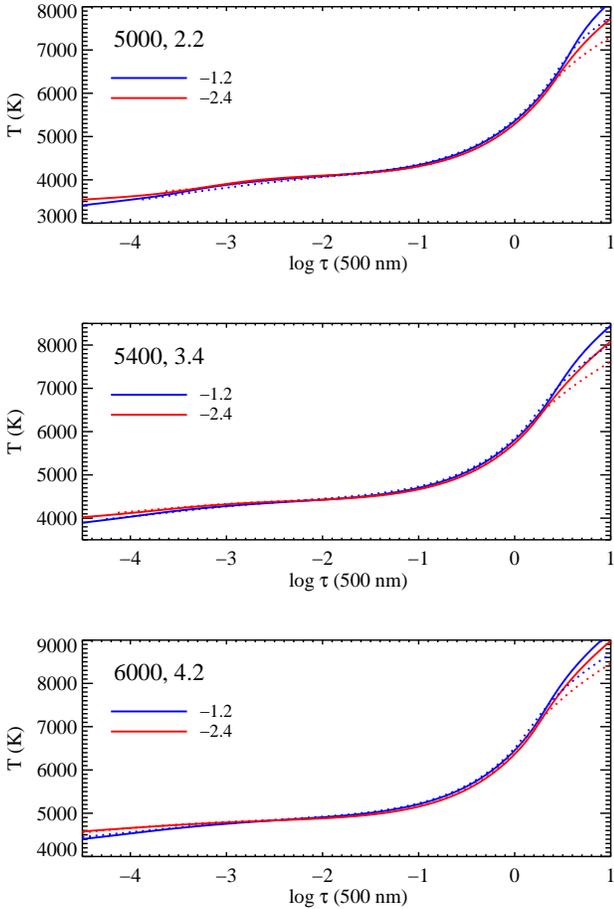}
\caption{Comparison of MARCS (dotted) and MAFAGS (solid) model atmospheres for
selected stellar parameters at two different metallicities ([Fe/H] $=
-1.2,-2.4$).}
\label{models}
\end{figure}

{\label{sec:atm}}
All calculations in this work were performed with classical 1D LTE
plane-parallel model atmospheres. We used the MAFAGS \citep{gruppa,gruppb} and
MARCS \citep{Gus08} models, which are both well-adapted to analysis of
late-type stars. The model atmosphere codes adopt slightly
  different prescriptions for the convective flux transport and background opacity. However,
comparison of the model T$(\tau)$ relation showed that these differences are
very small, and reveal themselves only in optically thick layers, where the
treatment of convection is important (see Fig. \ref{models}). 
A comparison of the MAFAGS and MARCS models was presented in \citet{MB1d3d},
where we showed that the differences in Fe abundances obtained with different
model atmosphere codes are very small, typically within $0.05$ dex. We
comment on this further in Sect. \ref{sec:abund}.

\subsection{Stellar parameters}{\label{sec:params}}

Stellar parameters for the selected sample of stars were taken from \cite[][`H']{hansen} and \cite[][ -- `BG';`B']{berggeh,MB1d3d}, giving preference to parameters determined with
IR photometry and parallaxes. The `H', `BG', and `B' indicate the source of the temperature hereafter. A brief description of these data follows.

The effective temperatures were derived from several different colour indices
and
calibration methods \citep{alondw,AlonsoG,ramirez,masana,Oenehag,casagra}. We
chose temperature calibrations that fall in the middle of the probed calibration
ranges \citep[see][for details]{hansen}. The reddening values were taken from
\citet{schlegel}, and since all the reddening values are much lower than
0.1\,mag \citep{bonEBV}, we have not applied their corrections to these
values. Our chosen stellar parameters are also consistent with the effective
temperatures ($\Teff$) determined from
the 1D fitting of Balmer profiles by \citet{gehren04,gehren06}. For the few stars (HD 19445,
HD 142038, G 64-12) we have in common with
 \citet{gehren04,gehren06}, they derive $\Teff = 5985$, $5773$, and
$6407$ K, respectively, with an \textit{rms} offset of 10 K from our values. This confirms the agreement between the Balmer $\Teff$ scale from \citet{gehren04,gehren06} and the method described in \citet{alondw} that we used here. Balmer line $\Teff$'s (1D) from the same reference agree with the values
we adopted for HD 134169,
HD 148816, and HD 184448. The parameters for HD 122563 and G 64-12 are
those from \citet{MB1d3d}.

For the stars with parallax measurements, the surface gravity was calculated
using the
classical formula that relates mass, temperature, magnitude and parallax to
gravity. Masses and bolometric corrections were taken from
\citet{nissengrav,nissen02,nissen07}. Metallicities were then initially
estimated in LTE, and the effects of NLTE were taken into account by applying
NLTE abundance corrections to Fe. These corrections were computed for the adopted \ion{Fe}{i}
line list by interpolation in the Fe NLTE grid presented by \cite{lindfe}. We
note that for stars with metallicity [Fe/H] $ > -2$, the systematic difference
between LTE abundances of \ion{Fe}{i} and \ion{Fe}{ii} is not large ($\leq$
0.1\,dex). The Sr NLTE corrections also stay within $0.05 - 0.07$ dex (see Sect. \ref{results}) consistent with the results in \citet{bergSr}. The effect
of NLTE becomes very important for Fe in very metal-poor stars with [Fe/H] $< -2.5$. In
particular, most of the stars from our comparison samples
\citep{francois,bonifaciodw} are subject to NLTE metallicity
corrections ranging from $+0.2$ to $+0.3$ dex. For the extremely
metal-poor (EMP; $-4 <$ [Fe/H] $<-3$) giant stars, we obtain a maximum NLTE correction 
  of $\Dm$ of $+0.3$ dex.
\begin{figure}[ht]
\centering
\includegraphics[width=0.5\textwidth]{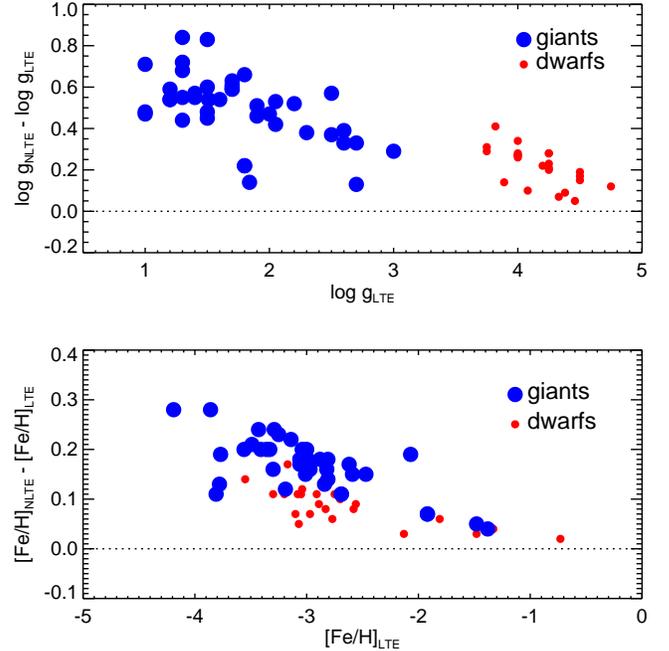}
\caption{Comparison of LTE and NLTE surface gravities (top) and metallicities
(bottom).}
\label{nlte_params}
\end{figure}

For the other stars, which do not have known parallaxes, the initial estimate of $\log g$
and [Fe/H] was obtained
from the LTE ionisation equilibrium of Fe, and Fe NLTE corrections were applied to
both gravity and metallicity. To compute the NLTE correction for surface gravity, we used the
approximate $\Dg$ - $\Dm$ calculations from \citet[][their Sect. 3]{lindfe}. The
changes in $\log g$ due to NLTE effects are significant and have a
clear effect on the [Sr/Fe] ratios derived from the gravity-sensitive \ion{Sr}{ii} lines, compared to
the changes in metallicity. The NLTE gravity corrections reach up to $\sim +0.8$
dex for the most metal-poor giants in our sample. The influence of $T_{\rm eff}$ on \ion{Sr}{ii} is minor. By changing the temperature with its uncertainty, the \ion{Sr}{ii} abundance changes by 0.01-0.05\,dex, while the change in \ion{Sr}{i} is much larger (see Sect. \ref{sec:uncertain}). The NLTE and LTE stellar
parameters are compared in Fig. \ref{nlte_params}.\\
\begin{figure}[!h]
\centering
\includegraphics[width=0.5\textwidth]{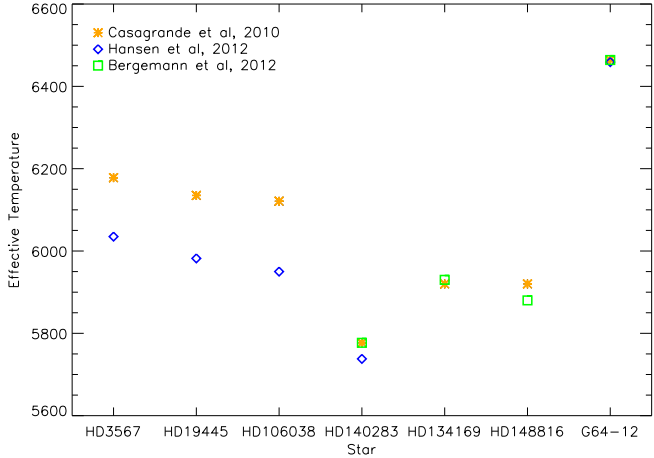}
\caption{Comparison of effective temperatures determined with different methods for seven stars. The legend indicates the original paper \citep[][ -- yellow '*', blue diamond, and green square, respectively]{casagra, hansen,bergSr} from which the temperatures have been taken.}
\label{tcompare}
\end{figure}

As seen from Fig. \ref{tcompare} the temperatures for seven stars taken from three different sources \citep{casagra,bergSr,hansen} agree within 40\,K in most cases, and for a few stars (e.g. HD106038) the difference between the two IRFM methods \citep{alondw,casagra} is 170\,K. This difference is within the combined errors, if we include systematic errors, as well as the uncertainty on $E(B-V)$.  The dwarfs from the comparison sample \citep{bonifaciodw} had their temperatures estimated from H$\alpha$ line profile fitting, which yielded values in good agreement with those determined from \citet{alondw} calibrations. We refer the reader to \citet{bonifaciodw} and \citet{sbordone} for further details on H$\alpha$ line profile fitting. The giants from the second comparison sample \citep{francois} had their temperatures determined by the use of broad range photometry calibrations from \citet{AlonsoG}. This is the same method we applied here, and the differences/offsets between our temperature determinations and those made for the comparison samples are minimal.

The microturbulence velocity was fixed by requiring that \ion{Fe}{i} lines yield
the same abundances regardless of their equivalent width. In our parameter
space, the NLTE corrections to microturbulence \citep{lindfe} are smaller than
the formal uncertainty of $\xi_t$ $(\pm0.15$ km/s) and were not considered here.

The main error in our photometric temperatures comes from reddening ($\lesssim
\pm$0.05\,mag). For Balmer lines, the temperature error is largely internal and
is determined from the profile fitting. This applies to the six stars marked by a `B' in the following. The error in the adopted $\log g$ values
is dominated by that of parallaxes ($\lesssim \pm$1.0"). The errors in [Fe/H]
and microturbulence are assumed to be $0.1(5)$\,dex and $0.15$\,km/s,
respectively. After propagating all stellar parameter uncertainties, we adopted
a common set of uncertainties for our stars of ($\Teff$/$\log g$/[Fe/H]/$\xi$):
$\pm$100\,K/0.2\,dex/0.1\,dex/0.15\,km/s. For the most metal-poor stars, we find
slightly higher values: $\pm$100\,K/0.25\,dex/0.15\,dex/0.15\,km/s. These errors
are internal to our method. \textit{The differences between LTE and NLTE stellar
parameters highlights that systematic errors in LTE are larger than the internal errors.}

\subsection{Sr abundance determinations}{\label{sec:abund}}
The NLTE statistical equilibrium calculations for Sr were performed with the
revised version of the DETAIL code \citep{butler85}. The new model atom of
Sr and other related aspects of the NLTE calculations are described in detail in
\citet{bergSr}. The NLTE effects on the \ion{Sr}{i} lines are primarily
caused by over-ionisation, which leads to systematically higher NLTE abundances
compared with LTE, especially for more metal-poor and hotter stars. In contrast,
deviations from LTE in the \ion{Sr}{ii} lines are largely driven by strong line
scattering. As a consequence, the differences between the LTE and NLTE
abundances may be positive or negative, depending on temperature, gravity, and
metallicity of a star.

The LTE and NLTE abundances of Sr in the selected metal-poor stars were
determined as
follows. 
The LTE abundances are synthesised with MOOG, while NLTE synthesised abundances
are derived using the SIU code
\citep{reetz}, where the NLTE departure coefficients are computed with DETAIL.
In both cases we apply the same atomic data. We note that the LTE
abundances from SIU and MOOG agree within 0.1\,dex. This difference is a
combination of local continuum placement (which can be up to 0.05\,dex when set locally by eye) and of the
different synthesis codes, since the model atmospheres are very similar, as
seen from Fig. \ref{models}, they almost do not contribute to this difference. Only very deep in the atmosphere (close to $\tau = 1$) do the MARCS and MAFAGS models differ due to a more efficient transport of
convective flux in the MARCS models, which leads to a cooler atmosphere than in MAFAGS.
For $18$ stars, with a four-star overlap with the SIU analysis, we first determine
LTE abundances by spectrum synthesis with the MOOG code. To obtain NLTE
abundances, we then apply the NLTE corrections calculated with DETAIL and MAFAGS. 
We do not perform differential abundance analysis with respect
to the Sun.
\begin{figure}
\centering
\includegraphics[width=0.48\textwidth]{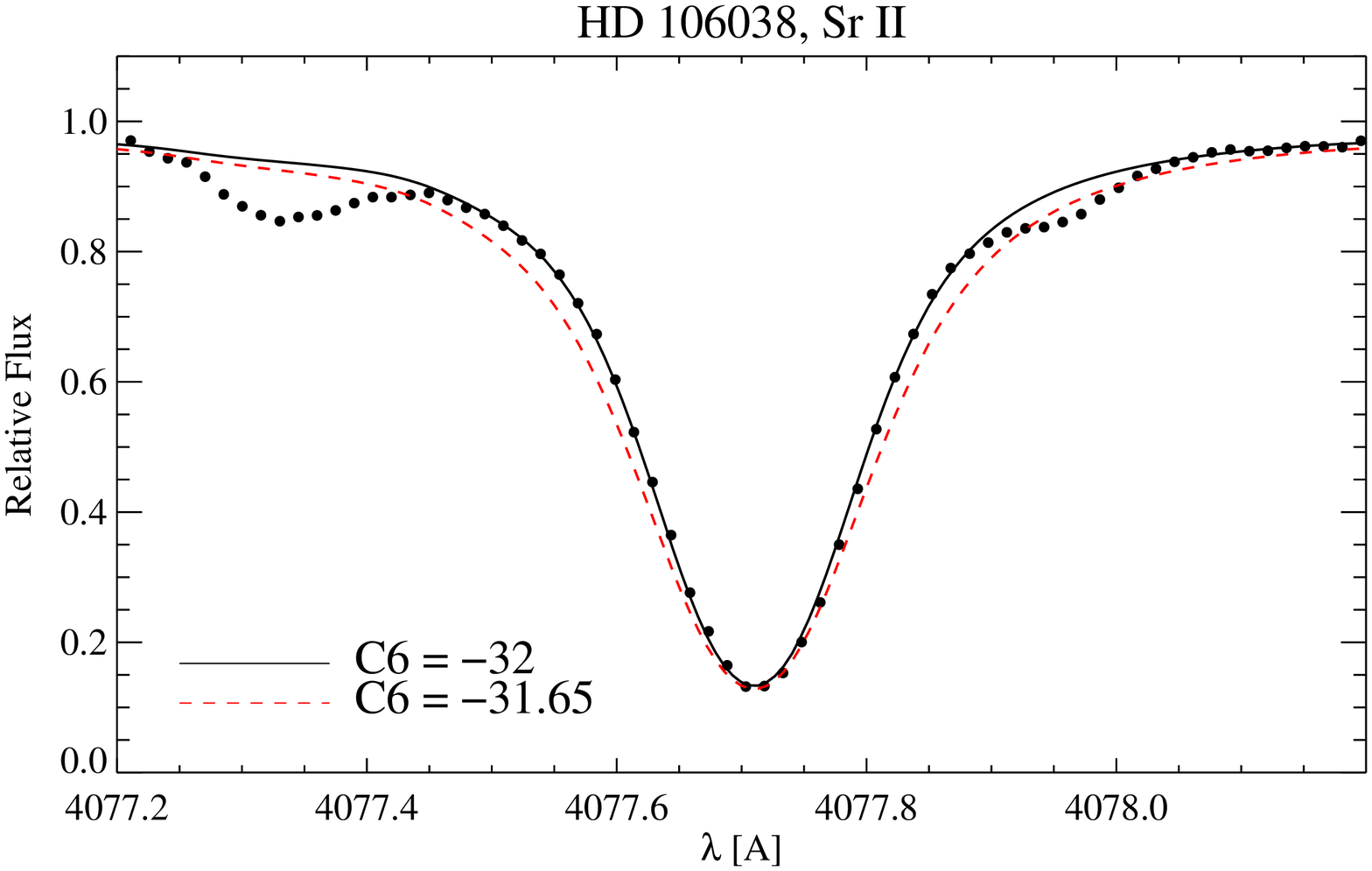}
\includegraphics[width=0.48\textwidth]{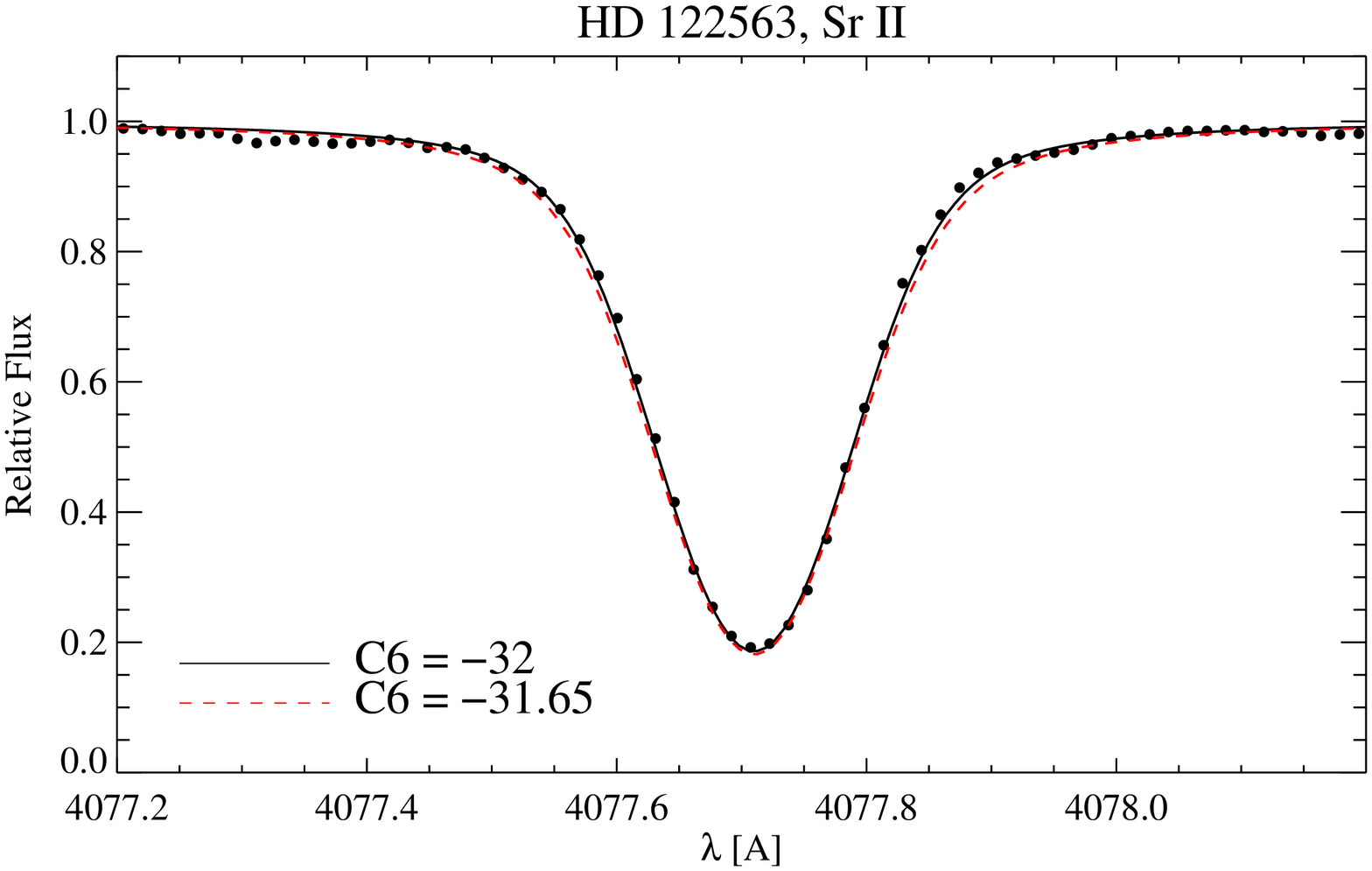}
\caption{Effect of elastic H I collisions on the 4077 \AA\, lines expressed through different C$_6$ values (red dashed and black solid line) - compared to observations (black dots) of a dwarf (HD 106038) and a metal-poor giant (HD 122563) star.}
\label{damping}
\end{figure}

\begin{figure}
\centering
\includegraphics[width=0.48\textwidth]{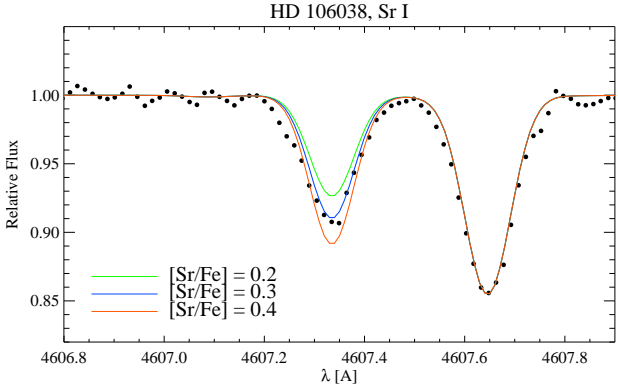}
\caption{Synthetic spectra (solid lines) of the 4607 \AA\, \ion{Sr}{i} line plotted on top of
the observed spectrum of HD 106038 (dots).}
\label{sr1}
\end{figure}

\begin{figure}[!h]
\centering
\includegraphics[width=0.48\textwidth]{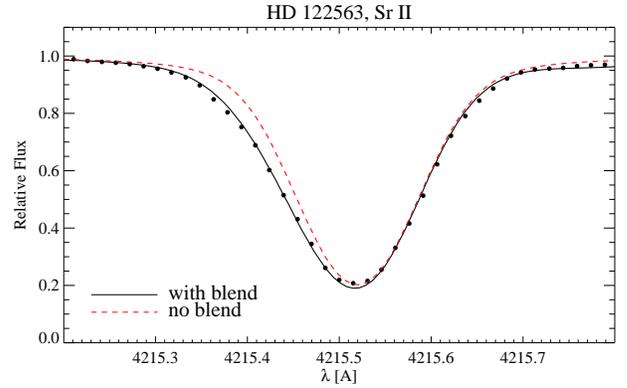}
\caption{Comparison of synthetic profiles (lines) of the \ion{Sr}{ii} line to observations of HD122563 (black dots). The syntheses have been computed
with (black solid line) and without (red dashed line) the \ion{Fe}{i} blends for the 4215\,\AA\, \ion{Sr}{ii} line.}
\label{blend}
\end{figure}

In particular, to remain consistent with our previous analysis, we use the van
der Waals broadening and $gf$-values from \cite{bergSr} (see Table \ref{gamma}). 
\begin{table}
\begin{center}
\caption[]{Atomic data for Sr I and II}
\label{gamma}
\begin{tabular}{l c c c c c }
\hline
\hline
Element & $\lambda$ & $\chi$ & log$gf$ & log($\gamma$/N$_H$) & log C$_6$\\
  & \AA & eV & & rad cm$^3$ s$^{-1}$ & cm$^6$ s$^{-1}$\\
\hline
Sr I & 4607.33 & 0.0 & 0.283 & -7.53 & -31.2\\
Sr II & 4077.71 & 0.0 & 0.158* & -7.81 & -32.0\\
\hline
\hline
\end{tabular}
\tablefoot{
\tablefoottext{*}{Total log gf value. Further details and hfs splitting can be
found in \cite{bergSr}.}
}
\end{center}
\end{table}
According to this study the damping constants for the \ion{Sr}{ii} lines could
be somewhat uncertain. Variation in $\log C_6$ by $\pm 0.35$ ($\sim 20 \%$)
leads to a change in abundances by $\mp 0.15$ dex for the 4077\AA\, \ion{Sr}{ii} line in the
metal-poor stars with [Fe/H]$ > -1.5$ (see Fig. \ref{damping}), but only has a minor effect at very low metallicity ($\sim \mp 0.05$ dex). 
The accuracy of the gf-value for the 4607 \AA\, \ion{Sr}{i} line was critically evaluated by
the
NIST database. The uncertainties are less than one percent and
yield accurate abundances even for metal-rich stars (see Fig. \ref{sr1}), whereas the
\ion{Sr}{ii} resonance lines at $4077$ and $4215$ \AA\ are too blended and
strong to give any reliable information about the solar Sr abundance (see Fig. \ref{blend}). We derive our \ion{Sr}{i} abundances from the 4607 \AA\, line and the \ion{Sr}{ii} abundances from the 4077 \AA\, line, since all our stars have sub-solar metallicities.

\begin{figure}
\centering
\includegraphics[width=0.49\textwidth]{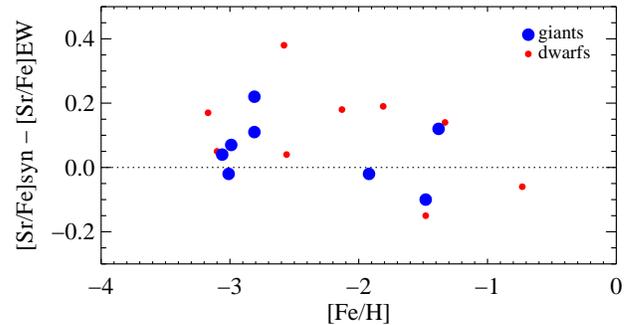}
\caption{Difference in Sr abundances between EW and synthesis. Dwarfs/giants are shown as small red/big blue filled circles, respectively.}
\label{diffewsyn}
\end{figure}

We tested the {\it Abfind} package in MOOG, which uses the measured equivalent
widths (EW) to compute abundance by the curve-of-growth method. We find,
however, that this method yields abundances in slight disagreement
($\pm0.04<$ [Sr/Fe]$_{\rm EW-syn} <\pm0.25$) with the results
obtained by the full profile fitting using the {\it Synth} (synthesis) package
of the same code (see the online material). Depending on the line properties, in
particular equivalent width, the abundance is either over- or
under-estimated (Table \ref{LTEabu}). This expected; the \ion{Sr}{i}
line is very weak in metal-poor stars, and both \ion{Sr}{ii} lines are strong, sensitive to damping, and affected by blends. For example, the EWs determined by
fitting Voigt and Gaussian profiles in IRAF ('/' separated entries in the 'EW'
columns in Table \ref{LTEabu}) differ generally by 5--10\% (in a few cases, like the VMP dwarf 
HD 106038, the difference is approximately a factor of three larger). As a result,
abundances derived using the EWs may be discrepant by up to $0.4$ dex (Fig.
\ref{diffewsyn}). However, except for one case (the EMP subgiant HD 140283), the largest
difference between EW and synthesis determined abundances are of the order of
$\pm0.1$ dex (see online material). This value might be slightly over-estimated due to local
continuum placement.

Blends influence abundance determinations in both the metal-poor and metal-rich
parts of our sample. Features in the blue and red wings of the $4077$ \ion{Sr}{ii} line, $\sim
0.3~\AA$ away from the line centre, only vanish in the spectra of most
metal-poor stars, [Fe/H] $< -2$. These blends are due to lanthanum, chromium, and
dysprosium. The
\ion{Sr}{ii} line at $4215$ \AA\ is blended by the two \ion{Fe}{i} lines at
$4215.42$ and $4216.18$ \AA. The former is rather strong and clearly distorts
the shape of the
\ion{Sr}{ii} profile, as directly seen in the very high-resolution spectra.
Figure \ref{blend} shows that even for the very metal-poor giant HD 122563 with
[Fe/H]$_{\rm NLTE} = -2.5$, the abundance is over-estimated by $0.14$ dex (when synthesised) if this
blend is not taken into account. Therefore, the abundance determined from the
4215 \AA\, \ion{Sr}{ii} line is subject to a systematic uncertainty, and the blend is 
sensitive to $T_{\rm eff}$, $\log g$, and metallicity.

On these grounds, we do not include the $4215$ \AA\ \ion{Sr}{ii} line in the
abundance calculations. Furthermore, for the comparison samples from
\cite{francois} and \cite{bonifaciodw}, only measurements for the 4077 \AA\, \ion{Sr}{ii} line are
available. To be consistent and to retain the full sample
size, we therefore only use the 4077 \ion{Sr}{ii} line.

\begin{table*}
\begin{center}
\caption[]{Example of LTE [Sr/Fe] abundances for a dwarf, sub-giant, and
giant from the measured equivalent widths (EW) and from synthesis (synt). The measured Gauss/Voigt EWs (G/V) are given in m\AA\,, together with the stellar parameters.}
\label{LTEabu}
\hspace{-2mm}
\begin{tabular}{l c c c c c c c c c c}
\hline
\hline
Star &  4077 EW & EW G/V& 4077 synt & 4607 EW & EW G/V& 4607 synt& T$_{\rm eff}$ & $\log g$ & [Fe/H]$_{LTE}$ & $\xi$ \\
   &  &  [m\AA] & & &  [m\AA] & & [K] & & & [km/s] \\
\hline
HD106038 &  $0.28/0.65$ & $186.1/271.0$ & 0.5 &$0.43/0.47$ & $13.1/14.1$ & 0.33  & 5950  & 4.33 & $-1.48$& 1.1\\  
HD140283 &  $-0.44/-0.33$ & $75.8/79.8$ & $-$0.15  & -- & -- & -- & 5777 & 3.70 & $-2.58$ & 1.5 \\     
HD122563 &   $-0.34/-0.09^*$ &$158.8/184.0$ & $-0.05$  &$-0.54^*$ &2.9 & $<-$0.6 & 4665 &1.65  & $-2.50$ &1.8\\  
\hline
\hline
\end{tabular}
\tablefoot{
\tablefoottext{*}{Value uncertain}
}
\end{center}
\end{table*}

The \ion{Sr}{i} line is generally weak, and regardless of the profile fitted to
this line, the EW-converted abundances tend to be larger than the synthesised
abundances. This might be due to an iron blend in the red wing of this neutral
strontium line (see Fig. \ref{sr1}, where a stronger Fe line blends into
the red wing of the \ion{Sr}{i} line).

The LTE abundances derived for our sample were calculated with LTE stellar
parameters (as described in Sect. \ref{sec:params}) using the 1D LTE synthetic
spectrum code MOOG to synthesise spectra for these stars. The EWs for the dwarf
comparison sample
were taken from \citet{bonifaciodw}. We measured the EWs for the remaining
stars. The LTE Sr II abundances for the comparison samples (dwarfs and giants)
were redetermined with MARCS models and MOOG, using the solar abundance,
from \cite{anders}, which we have adopted for this study. The LTE abundances we
calculated agree within 0.05 - 0.1 dex with those published in \cite[][F07]{francois}
and \cite[][B09]{bonifaciodw}, and only for a handful of stars is the difference
greater than 0.2 dex.

The correction for NLTE effects is most conveniently performed by
differentiating LTE and NLTE curves-of-growth at a given line strength. The
method we adopted thus relies on the determination of EWs and subsequent
translation to LTE and NLTE abundances. 
From the comparison sample (the {\it ``First Stars"} samples), we only have these EWs.
When determining EWs by profile fitting with a single component, unresolved
blends may play a role, so we derived both LTE and NLTE EW-based abundances and compared them to synthesised
abundances to assess the impact the single-line assumption have on the
final abundances. For all the tests related to atomic data, blends and profile fitting, we have maintained one set of stellar parameters for each star (those listed in Table \ref{sample}). Thus, the resulting difference in abundance is an expression of uncertainties in the atomic data, unknown blends, and continuum placement. 

From Fig. \ref{diffewsyn} we estimate that the Sr abundances from the metal-rich
stars might be overestimated when using the EW method, while Sr in the metal-poor stars will be
overestimated with $< 0.32$ dex.
Another part of the over-/under-estimation can be assigned to the line profile
fitting and continuum placement.

\section{Results: LTE vs NLTE}{\label{results}}

\begin{figure}[h!]
\centering
\includegraphics[width=0.5\textwidth]{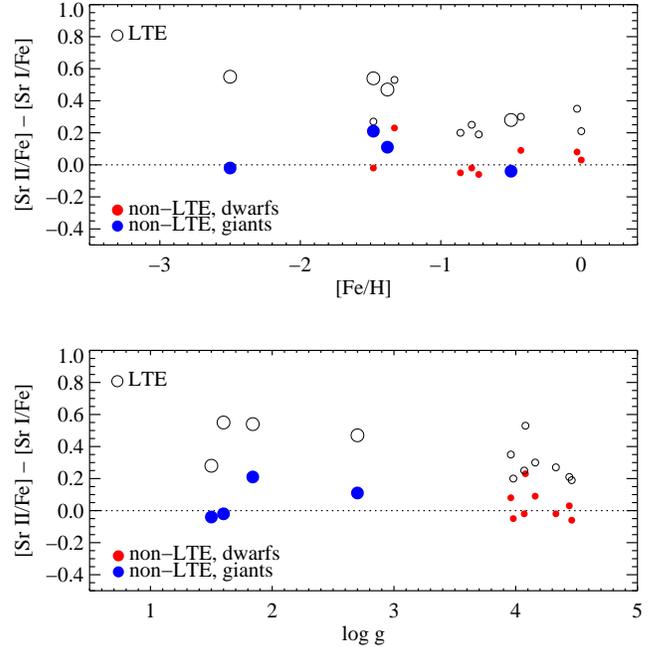}
\caption{Sr LTE (open circles) and NLTE abundance differences (filled big blue/small red circles for giants/dwarfs) as a function of [Fe/H] (top panel) and
$\log g$ (bottom panel).}
\label{nlte}
\end{figure}

The results for our sample are summarised in Table \ref{sample}, which provides
the mean of synthesised 
abundances for \ion{Sr}{i} and \ion{Sr}{ii}. Their differences are shown in
Fig. \ref{nlte} as a function of stellar [Fe/H] and $\log g$. The LTE and NLTE
line by line abundances from both EW and synthesis are also given in the online Table
\ref{allabun} (for our sample).
\begin{table*}
\begin{center}
\hspace{-4mm}
\caption{Stellar parameters and derived Sr abundances $\pm$ standard
deviation for the selected sample of stars. Strontium abundances with superscript (1) are derived only from the 4077\AA\, \ion{Sr}{ii} line. Parameters with subscript have been corrected for NLTE effects. Stars with an `s' superscript have had their abundances derived in SIU. \label{sample}}
\begin{tabular}{l c c c c c c c r r c}
\hline
\hline
  Star & $\pi \pm \sigma$ & T$_{\rm eff}$ & $\log g$ &  $\log g_{corr.}$ & [Fe/H]$_{LTE}$ & [Fe/H]$_{NLTE}$
& $\xi $ & [Sr/Fe]$_{LTE}$ & [Sr/Fe]$_{NLTE}$ & Comment\\
 & [mas]&[K] & & & & & km/s& & & \\ 
\hline
\object{HD 3567}         &  9.57$\pm$1.38 & 6035  & 4.08 & 4.08   & $-1.33$ & $-1.29$ & 1.5  &  $-0.03\pm0.8^x$ &  $0.06\pm0.8^x$ & [u,H]\\
\object{HD 19445}        & 25.85$\pm$1.14 & 5982  & 4.38 & 4.38   & $-2.13$ & $-2.10$ & 1.4 &  $0.13\pm0.8^x$  & $0.16\pm0.8^x$ & [u,H]\\
\object{HD 106038}       &  9.16$\pm$1.50 & 5950  & 4.33 & 4.33   & $-1.48$ & $-1.45$ & 1.1 &  $0.42\pm$0.12  & $0.45\pm0.21$   & [u,H]\\
\object{HD 121004}       & 16.73$\pm$1.35 & 5711  & 4.46 & 4.46   & $-0.73$ & $-0.71$ & 0.7 &  $0.18\pm0.04$  & $0.26\pm0.17$   & [u,H]\\
\object{HD 122196}       &  9.77$\pm$1.32 & 6048  & 3.89 & 3.89   & $-1.81$ & $-1.75$ & 1.2  & $0.24^{(1)}$     & $0.19^{(1)}$          & [u,H]\\
\object{HD 134169}       & 16.80$\pm$1.11 & 5930  & 3.98 & 3.98   &  \nod   & $-0.86$ & 1.8 & $-0.05\pm0.14$ & $-0.06\pm0.13$& [f,s,BG] \\
\object{HD 140283}       & 17.16$\pm$0.68 & 5777  & 3.70 & 3.70   & $-2.58$ & $-2.38$ & 1.5 &  $-0.15^{(1)}$ & $-0.37^{(1)}$  & [u,H,B] \\
\object{HD 148816}       & 24.34$\pm$0.90 & 5880  & 4.07 & 4.07   &  \nod   & $-0.78$ & 1.2 &  $-0.13\pm0.18$ & $-0.13\pm0.17$& [f,s,BG]\\
\object{HD 184448}       & 19.16$\pm$0.63 & 5765  & 4.16 & 4.16   &  \nod   & $-0.43$ & 1.2 &  $0.00\pm0.21$  & $-0.01\pm0.21$& [f,s,BG]\\
\object{G 64-12}         &  0.57$\pm$2.83 & 6464  & 4.30 &  4.30  & $-$3.24 & $-3.12$ & 1.5 &  $0.00^{(1)}$ & $0.17^{(1)}$           & [u,B] \\
\object{G 64-37}         &  2.88$\pm$3.10 & 6494  & 3.82$^*$ & $4.23$ & $-3.17$ & $-3.00$ & 1.4 &  $0.08^{(1)}$ & $0.17^{(1)}$            & [u,H] \\
\object{HD 122563}       &  4.22$\pm$0.35 & 4665  & 1.65 & 1.65   & $-$2.60 & $-2.50$ & 1.8 & $-0.23\pm0.8^x$ & $-0.12\pm0.8^x$ & [u,B] \\
\object{HD 175305}       &  6.18$\pm$0.56 & 5100  & 2.70 & 2.70  & $-1.38$ & $-1.34$ & 1.2 & $-0.12\pm0.32$ & $0.04\pm0.1$            & [h,H]\\
\hline
\object{BD -133442}      &     --         & 6450  & 4.20 & $4.42$ & $-2.56$ & $-2.47$ & 1.5 & $0.30^{(1)}$ & $0.21^{(1)}$          & [u,H] \\
\object{CS 30312-059}    &     --         & 5021  & 1.90 & $2.41$ & $-3.06$ & $-2.89$ & 1.5 & $0.50^{(1)}$ & $0.31^{(1)}$          & [h,H]   \\
\object{CS 31082-001}    &     --         & 4925  & 1.51 & $2.05$ & $-2.81$ & $-2.63$ & 1.4 & $0.70^{(1)}$ & $0.60^{(1)}$          & [h,H] \\
\object{HD 74462}        &     --         & 4590  & 1.84 & $1.98$ & $-1.48$ & $-1.43$ & 1.1 & $-0.25\pm0.35$ & $-0.14\pm0.03$  & [h,H] \\
\object{HD 126238}       &     --         & 4900  & 1.80 &  2.02  & $-1.92$ & $-1.85$ & 1.5 & $-0.17\pm0.24$& $0.01\pm0.06$    & [u,H]  \\ 
\object{HD 126587}       &     --         & 4950  & 1.90 & $2.36$ & $-3.01$ & $-2.86$ & 1.65& $0.23\pm0.8^x$ & $0.22\pm0.8^x$  & [u,H] \\
\object{HE 0315+0000}    &     --         & 5050  & 2.05 & $2.47$ & $-2.81$ & $-2.67$ & 1.7 & $0.39^{(1)}$ & $0.23^{(1)}$          & [u,H] \\
\object{HE 1219-0312}    &     --         & 5100  & 2.05 & $2.58$ & $-2.99$ & $-2.81$ & 1.65& $0.29^{(1)}$ & $0.12^{(1)}$          & [u,H]\\
\hline
\hline
\end{tabular}
\tablefoot{
\tablefoottext{*}{Value from ionisation equilibrium}\\
\tablefoottext{B, BG, H}{Temperature from \cite[][B]{MB1d3d}, \cite[][BG]{berggeh}, and \cite[][H]{hansen}, respectively.}\\
\tablefoottext{u,f,h}{Observed spectra from$:$ $^u$ UVES/VLT, $^f$
FOCES/Calar-Alto \cite{gehren04,gehren06}, $^h$ HIRES/Keck}\\
\tablefoottext{x}{Weighted average, which includes an upper limit that was given
half weight. The large uncertainty is a reciprocal square root of the summed
weights.}\\
}
\end{center}
\end{table*}

The LTE approximation fails to establish the
ionisation balance of \ion{Sr}{i} and \ion{Sr}{ii}. Figure \ref{nlte} shows that
the offset between the two ionisation stages is about $0.2$ dex for dwarfs, but
it increases up to $0.5$ dex for giants. This difference is mainly caused by the
progressively increasing systematic error in the LTE abundance inferred from
the \ion{Sr}{i} line, which shows NLTE abundance corrections of up to $0.5$
dex at low metallicity and low gravity (online Table \ref{allabun}).
Although the NLTE effects on the resonance \ion{Sr}{ii} lines are not significantly
pronounced, they depend on stellar parameters, particularly on the [Fe/H] or, equivalently, on the
Sr abundance itself (see discussion in Sect. \ref{sec:abund}). In Table
\ref{sample}, we see that the LTE abundances obtained from the \ion{Sr}{ii} lines can be over- or
under-estimated by up to $0.1$ dex, which may introduce a
spurious systematic trend or, more likely give rise to a larger
line-to-line scatter. If the gravities are derived from parallaxes, their values will be the same in both LTE and NLTE; however, when derived from ionisation equilibrium, the NLTE corrected $\log g_{corr}$ will differ from the LTE \ion{Fe}{i} based $\log g$. For our stellar sample, even though the
  line-to-line scatter is clearly smaller under NLTE, the star-to-star
scatter almost remains the same under both LTE and NLTE. The results for the comparison samples (the {\it `First Stars'} samples -- F07, and B09) are shown in Table \ref{compsample}.

\begin{table*}[!ht]
\begin{center}
\caption{Basic parameters and EW calculated Sr abundances from the 4077\AA\,
line for the comparison stars. Top part contains dwarf stars from the comparison sample, while lower part shows giants from the second comparison sample. Details and subscripts are described in Table \ref{sample}.}\label{compsample}
\begin{tabular}{l c c c c c c r r}
\hline
\hline
Star &  T$_{\rm eff}$ & $\log g$ & $\log g_{corrected}$ & [Fe/H]$_{LTE}$ & [Fe/H]$_{NLTE}$ & $\xi $ &
[Sr/Fe]$_{LTE}$ & [Sr/Fe]$_{NLTE}$\\
  & [K] & & & & & [km/s] &  & \\
\hline
Dwarf sample$^{FS, B09}$  & & & & & & & & \\
   \object{BS16023-046}   & 6364  &  4.50  &  4.69  & $-$2.97  & $-$2.90  &  1.3  &  $-0.20$ 
& $-0.17$\\    
    \object{BS16076-006}  &  5199 &   3.00 &   3.29 &  $-$3.81 &  $-$3.70 &  1.4  &  0.67   
& -0.64\\
    \object{BS16968-061}  &  6035 &   3.75 &   4.04 &  $-$3.05 &  $-$2.94 &  1.5  & $-1.59$ 
& $-1.58$\\
    \object{BS17570-063}  &  6242 &   4.75 &   4.87 &  $-$2.92 &  $-$2.87 &  0.5  &  0.03   
& $-0.02$\\
    \object{CS22177-009}  &  6257 &   4.50 &   4.67 &  $-$3.10 &  $-$3.03 &  1.2  &  $-0.15$
& $-0.12$\\
    \object{CS22888-031}  &  6151 &   5.00 &   5.09 &  $-$3.30 &  $-$3.26 &  0.5&    0.05   
&  0.09\\
    \object{CS22948-093}  &  6356 &   4.25 &   4.53 &  $-$3.30 &  $-$3.19 &  1.2&   $-0.08$ 
&  $-0.01$\\
    \object{CS22953-037}  &  6364 &   4.25 &   4.48 &  $-$2.89 &  $-$2.80 &  1.4&   $-0.45$ 
&  $-0.46$\\
    \object{CS22965-054}  &  6089 &   3.75 &   4.06 &  $-$3.04 &  $-$2.92 &  1.4&   $-2.09$ 
&  $-2.17$\\
    \object{CS22966-011}  &  6204 &   4.75 &   4.87 &  $-$3.07 &  $-$3.02 &  1.1&    0.88   
&  0.95\\
    \object{CS29499-060}  &  6318 &   4.00 &   4.26 &  $-$2.70 &  $-$2.60 &  1.5&    $-0.63$
&  $-0.70$\\
    \object{CS29506-007}  &  6273 &   4.00 &   4.27 &  $-$2.91 &  $-$2.80 &  1.7&    $-0.49$
&  $-0.50$\\
    \object{CS29506-090}  &  6303 &   4.25 &   4.46 &  $-$2.83 &  $-$2.75 &  1.4&    0.33   
&  0.27\\
    \object{CS29518-020}  &  6242 &   4.50 &   4.65 &  $-$2.77 &  $-$2.71 &  1.7&    0.08   
&  0.05\\
    \object{CS29518-043}  &  6432 &   4.25 &   4.53 &  $-$3.20 &  $-$3.09 &  1.3 &   \nod      
& \nod\\
    \object{CS29527-015}  &  6242 &   4.00 &   4.34 &  $-$3.55 &  $-$3.41 &  1.6 &   0.12   
& 0.22\\
    \object{CS30301-024}  &  6334 &   4.00 &   4.27 &  $-$2.75 &  $-$2.64 &  1.6 &   $-0.41$
& $-0.47$\\
    \object{CS30339-069}  &  6242 &   4.00 &   4.28 &  $-$3.08 &  $-$2.97 &  1.3 &   0.43   
& 0.40\\
    \object{CS31061-032}  &  6409 &   4.25 &   4.45 &  $-$2.58 &  $-$2.50 &  1.4 &   $-0.48$
& $-0.54$\\
  \hline
Giant  sample$^{FS, F07}$ & & & & & & & & \\
 \object{BD+17:3248}   &  5250 &   1.40  &  1.97 &  $-$2.07  & $-$1.88 &  1.5 &    0.00     &
$-0.09$\\
\object{BD-18:5550}    &  4750 &   1.40  &  1.95 &  $-$3.06  & $-$2.88 &  1.8 &   $-1.03$  &
$-1.12$\\
\object{BS16467-062}   &  5200 &   2.50  &  3.07 &  $-$3.77  & $-$3.58 &  1.6 &   $-1.96$  &
$-1.99$\\
\object{BS16477-003}   &  4900 &   1.70  &  2.29 &  $-$3.36  & $-$3.16 &  1.8 &    $0.06$  &
$-0.19$\\ 
\object{BS17569-049}   &  4700 &   1.20  &  1.74 &  $-$2.88  & $-$2.70 &  1.9 &    0.17    &
$-0.18$\\ 
\object{CD-38:245}     &  4800 &   1.50  &  2.33 &  $-$4.19  & $-$3.91 &  2.2 &    $-0.72$ &
$-0.82$ \\
\object{CS22169-035}   &  4700 &   1.20  &  1.79 &  $-$3.04  & $-$2.84 &  2.2 &    -0.33   &
$-0.30$\\ 
\object{CS22172-002}   &  4800 &   1.30  &  2.14 &  $-$3.86  & $-$3.58 &  2.2 &   $-1.40$  &
$-1.60$\\
\object{CS22186-025}   &  4900 &   1.50  &  2.10 &  $-$3.00  & $-$2.80 &  2.0 &    0.61    &
0.41\\
\object{CS22189-009}   &  4900 &   1.70  &  2.33 &  $-$3.49  & $-$3.28 &  1.9 &  $-1.01$   &
$-1.07$\\  
\object{CS22873-055}   &  4550 &   1.00  &  1.48 &  $-$2.99  & $-$2.83 &  2.2 &  $-0.16$   &
$-0.30$ \\
\object{CS22873-166}   &  4550 &   1.00  &  1.47 &  $-$2.97  & $-$2.81 &  2.1 &     0.07   &
$-0.07$\\
\object{CS22878-101}   &  4800 &   1.30  &  1.98 &  $-$3.25  & $-$3.02 &  2.0 &  $-0.41$   &
$-0.36$\\
\object{CS22885-096}   &  5050 &   2.60  &  2.99 &  $-$3.78  & $-$3.65 &  1.8 &  $-1.44$  &
$-1.47$\\
\object{CS22891-209}   &  4700 &   1.00  &  1.71 &  $-$3.29  & $-$3.05 &  2.1 &    0.15   &
$-0.01$\\ 
\object{CS22892-052}   &  4850 &   1.60  &  2.14 &  $-$3.03  & $-$2.85 &  1.9 &    0.41   &
0.20\\
\object{CS22896-154}   &  5250 &   2.70  &  3.03 &  $-$2.69  & $-$2.58 &  1.2 &    0.32   &
0.18\\
\object{CS22897-008}   &  4900 &   1.70  &  2.30 &  $-$3.41  & $-$3.21 &  2.0 &  $-0.39$  &
$-1.20$\\
\object{CS22948-066}   &  5100 &   1.80  &  2.46 &  $-$3.14  & $-$2.92 &  2.0 &    0.46   &
0.25\\  
\object{CS22952-015}   & 4800  &  1.30  &  2.02  & $-$3.43  & $-$3.19 &   2.1 &  $-0.96$  &
$-1.07$ \\
\object{CS22953-003}   & 5100  &  2.30  &  2.68  & $-$2.84  & $-$2.71 &   1.7 &    0.13   &
$-0.01$\\
\object{CS22956-050}   &  4900 &   1.70 &   2.29 &  $-$3.33 &  $-$3.13&   1.8 &   $-0.47$ &
$-0.42$\\  
\object{CS22966-057}   & 5300  &  2.20  &  2.72  & $-$2.62  & $-$2.45 &   1.4 &   $-0.33$ &
$-0.44$\\
\object{CS22968-014}   & 4850  &  1.70  &  2.31  & $-$3.56  & $-$3.36 &   1.9 &   $-1.81$ &
$-1.69$\\
\object{CS29491-053}   & 4700  &  1.30  &  1.85  & $-$3.04  & $-$2.86 &   2.0 &   $-0.15$   &
$-0.36$\\
\object{CS29495-041}   & 4800  &  1.50  &  1.98  & $-$2.82  & $-$2.66 &   1.8 &   $-0.23$ &
$-0.36$\\
\object{CS29502-042}   & 5100  &  2.50  &  2.87  & $-$3.19  & $-$3.07 &   1.5 &   $-1.99$ &
$-2.05$\\
\object{CS29516-024}   & 4650  &  1.20  &  1.74  & $-$3.06  & $-$2.88 &   1.7 &  $-0.47$   &
$-0.61$\\
\object{CS29518-051}   & 5200  &  2.60  &  2.93  & $-$2.69  & $-$2.58 &   1.4 &    0.06   &
$-0.07$\\
\object{CS30325-094}   & 4950  &  2.00  &  2.47  & $-$3.30  & $-$3.14 &   1.5 &  $-2.38$  &
$-2.47$\\
\object{HD2796}        & 4950  &  1.50  &  1.95  & $-$2.47  & $-$2.32 &   2.1 &  $-0.23$   &
$-0.24$\\
\object{HD186478}      & 4700  &  1.30  &  1.74  & $-$2.59  & $-$2.44 &   2.0 &     0.05   &
$-0.12$\\
\hline
\hline
\end{tabular}
\tablefoot{
\tablefoottext{FS}{The synthesised LTE abundances can be found in the {\it First
Stars} papers; Dwarfs from \citet[][B09]{bonifaciodw}, and giants from \citet[][F07]{francois}}\\
}
\end{center}
\end{table*}

\begin{figure}[!ht]
\centering
\includegraphics[width=0.5\textwidth]{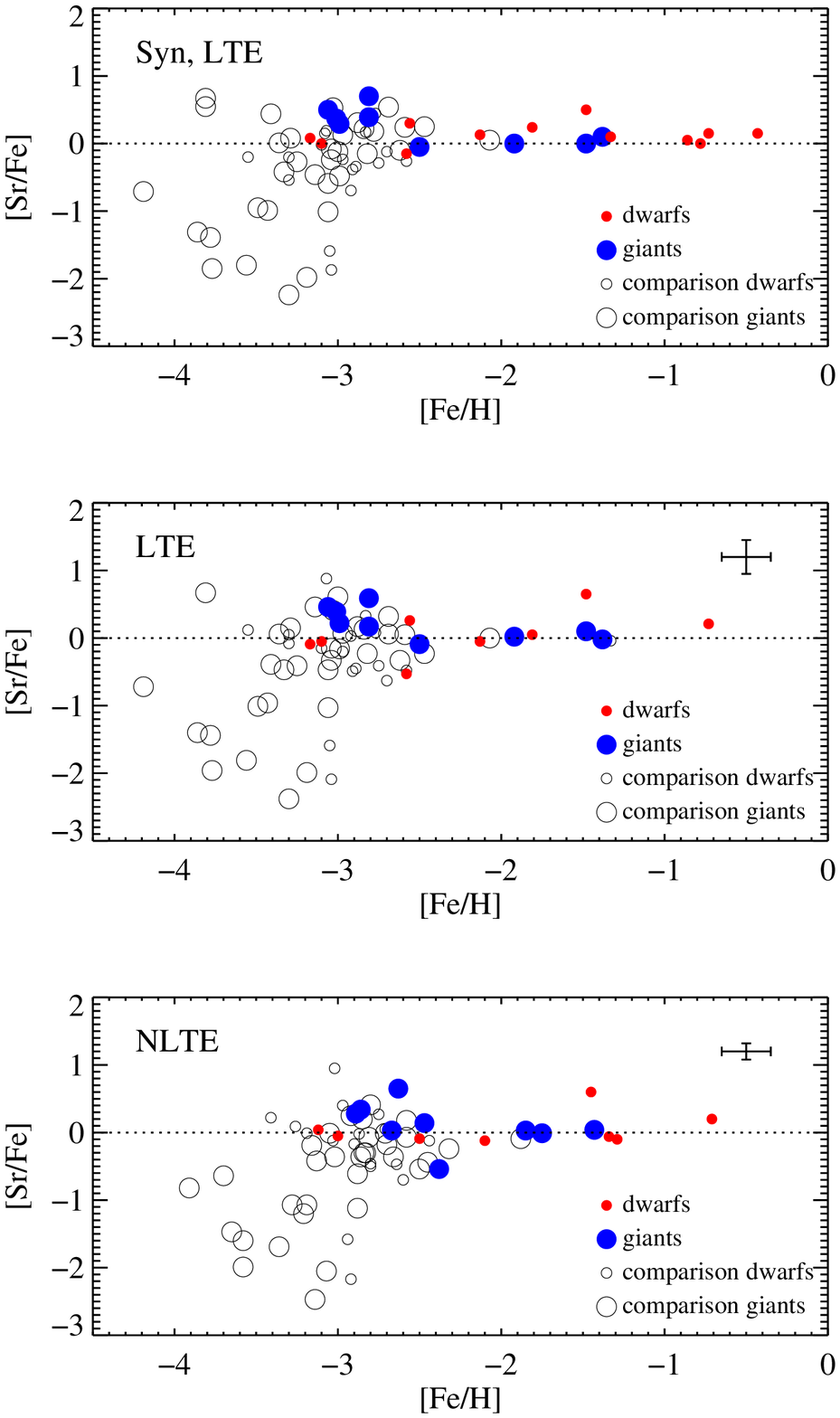}
\caption{Upper figure: LTE Sr synthesised (Syn) abundances for F07, B09 (open
circles), and our sample (filled
circles - giants big, blue; dwarfs small, red). The figure in the middle shows the EW-based abundances for the same
stars. Lower figure: The same samples but with NLTE corrected stellar
parameters, and NLTE corrected EW-based Sr abundances.}
\label{Srfig}
\end{figure}

Figure \ref{Srfig} shows the [Sr/Fe] ratios as a function of [Fe/H]. In the
figure, the error bars are the total, propagated uncertainties,
computed as described below. The exceptions are \object{HD 3567}, \object{HD 19445}, \object{HD 122563}, and {\object{HD 126587} with upper limits on the
abundance from the $4607$ \AA\ \ion{Sr}{i} line.

\begin{figure}
\centering
\includegraphics[width=0.31\textwidth, angle=-90]{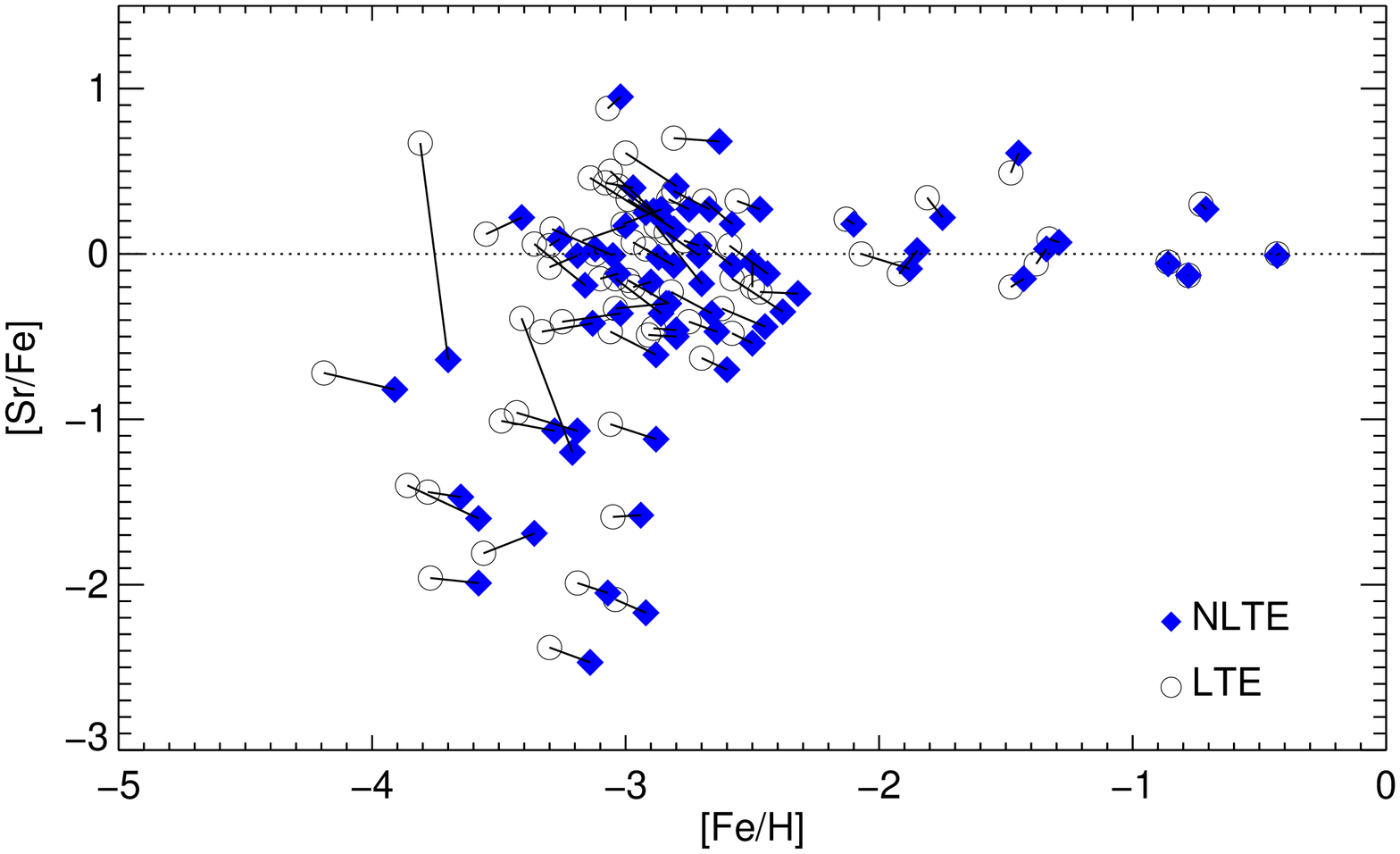}
\caption{Difference in [Sr/Fe] ratios between NLTE (blue diamonds) and LTE
(open circles).}
\label{ltenltefig}
\end{figure}

\subsection{Uncertainties}{\label{sec:uncertain}}
A number of test calculations varying the input parameters in the spectrum
synthesis were performed for the two representative stars with the same
metallicity: HD
106038 (dwarf) and HD 74462 (giant). In particular, we are interested in the
sensitivity of the abundances to the model atmosphere parameters: $\Teff$,
$\log g$, [Fe/H], and microturbulence (Table \ref{uncert}). As seen from this
table, the \ion{Sr}{ii} line at 4077 $\AA$ is very strong and mainly sensitive
to $\log g$ and $\xi$. In comparison, the neutral Sr abundance from the weaker 4607
\AA\, line is sensitive to temperature and [Fe/H], but almost not
affected by $\log g$ and $\xi$.

Several of the \ion{Sr}{i} abundances presented in the online Table \ref{allabun} are only
upper limits, which is why we applied a weighted average and a corresponding
reciprocal square root of the summed weights \citep[see][]{taylor}. The final
[Sr/Fe] abundances and their uncertainties are shown in Table \ref{sample}. The
upper limits of \ion{Sr}{i} were given half the weight of \ion{Sr}{ii}. This
overestimates the uncertainty a bit. However, with only two measurements, this
approach seems sensible. In some stars only \ion{Sr}{ii} could be measured,
so the abundance is based on only one trustworthy line. To estimate the
uncertainty in this case we made independent measurements of the 4077 \ion{Sr}{ii}
line, and found the derived abundances to be consistent to within 0.0 - 0.1 dex.
An average value of $\pm 0.05$ dex was adopted instead of the standard deviation
otherwise applied to the stars with two detectable lines.
\begin{table}[!h]
\begin{center}
\caption{Uncertainties in the individual LTE Sr I and II abundances in a giant
(\object{HD 74462}) and a dwarf (\object{HD 106038}) star.}
\label{uncert}
\begin{tabular}{l c c }
\hline
\hline
HD 74462: $[$Sr/Fe$]$ & 0.1 & -0.5  \\
Parameter/line [\AA]    & 4077 & 4607 \\
\hline
T $\pm100$         & $\pm0.03$ & $\pm0.17$ \\
logg $\pm0.2$      & $\pm0.04$ & $\pm0.01$\\
$[$Fe/H$] \pm0.1$ & $\pm0.12$ & $\pm0.09$\\
$\xi \pm0.15$       & $\pm0.07$ & $\pm0.01$\\
\hline
Propagated uncertainty & $\pm0.15$ & $\pm0.19$ \\
\hline
HD 106038: [Sr/Fe]    & 0.65   &  $0.33$ \\
T $\pm100$         & $\pm0.04$   &  $\pm0.09$  \\
logg $\pm0.2$      & $\pm0.17$   &   $\pm0.01$  \\    
$[$Fe/H$] \pm0.1$ & $\pm0.22$ &    $\pm0.09$  \\
$\xi \pm0.15$       & $\pm0.13$  &    $\pm0.01$  \\
\hline
Propagated uncertainty &  $\pm0.3$ & $\pm0.13$ \\
\hline
\hline
\end{tabular}
\end{center}
\end{table}

\textit{We summarise that the assumption of LTE especially in metal-poor, low-gravity stars will, in addition to not
fulfilling Sr ionisation balance, also introduce a weak
spurious trend of Sr abundances with metallicity.} The \ion{Sr}{i} resonance line
would consistently under-estimate the Sr abundance trend, whereas the
abundances obtained from the subordinate \ion{Sr}{ii} lines would be
systematically too large by $\sim 0.05$ dex. 
Whereas the abundances derived from the \ion{Sr}{i} line are significantly
affected by NLTE line formation, \ion{Sr}{ii} lines are less so. However, NLTE
effects for Fe must be accounted for in the spectroscopic gravity determination
in order to derive accurate abundances for Sr II lines.
{\it Furthermore, we note that the large star-to-star scatter found in LTE
abundance studies remains under NLTE even at extremely low metallicities.}

Our analysis of the chemical evolution of Sr in NLTE is different from previous
studies. \cite{andrievsky09} has already performed NLTE calculations for Sr;
however, their stellar parameters were determined assuming LTE. These have a
measurable impact on the Sr abundances. 
The difference in the overall trend of [Sr/Fe] with metallicity (\citet{bergSr}
and our Fig. \ref{Srfig}) is detectable, and is best seen in Fig.
\ref{ltenltefig}. As discussed in \citet{bergSr}, the model atom by \cite{andrSr}
is less complete than ours, and we incorporated new atomic data, which
influences the magnitude of NLTE abundance corrections. For the resonance
\ion{Sr}{ii} line at 4077 \AA, our NLTE corrections are mildly negative for any
$\log g$ and T$_{eff}$ at [Fe/H] $= -3$, whereas \cite{andrSr} obtain large positive
corrections for dwarfs and negative $\Delta$NLTE for giants.

\section{Discussion --- Chemical evolution of Sr}
\label{sec:chem}
Since we wish to assess the impact of the LTE assumption vs NLTE on the chemical
evolution of Sr, we selected a handful of Sr yields, covering both s- and
r-process contributions. We probed how LTE vs NLTE abundances behave in a Galactic chemical
evolution scheme. The yields will briefly be outlined below.

\subsection{Theoretical predictions of stellar Sr yields}
Here we consider the weak r-process yields from \citet{arcones} and
\citet{wanLet}, and the s-process yields from \citet{bis2010}, \citet{frisch},
and our own AGB yields based on the calculations presented in \citet{karakas}
and \citet{lugaro} that extend down to low metallicities of [Fe/H] = -2.3.

The yields from \citet{wanLet} describe the ejecta from low-mass
($\sim9M_{\odot}$) faint core-collapse electron-capture supernovae (ECSN). These may
occur frequently even at low metallicity \citep[e.g.][]{langer}, and their yields can therefore not be
neglected when considering the evolution of Sr. The amount of Sr injected into
the interstellar medium (ISM) is,
based on the self-consistent two-dimensional ECSN model of \citet{wanLet}, 1.79 $\cdot 10^{-4}
M_{\odot}$, which is obtained from the ejecta with an entropy
ranging from 10 to 25\,k$_B$/baryon and
an electron fraction (Y$_e$) between $0.4$ and $0.56$. This corresponds to the standard
yield ($W_{stand}$). 
Low $Y_e$ values are expected in low-mass progenitors due to their fast
explosions. Convective
bubbles expand fast enough to inhibit the neutrino reaction that increases the electron
fraction.
However, owing to the self-consistency of the explosion model there are no free parameters in the simulations once the progenitor model and physics input are chosen. 
The uncertainties therefore stem from the systematic uncertainties associated with the input to the SN
model and possible resolution limitations, but mainly from the progenitor mass range we adopt in
the GCE models, which is what gives rise to Wlow and Whigh. Here we have
selected two mass intervals for Wlow and Whigh that are representative of the SN
mass used for these SN model calculations (see also \citet{langer,nomoto87}).

Neutrino-driven winds following
immediately after supernova explosions, also from more massive progenitors,
will likewise contribute to the
amount of Sr in the ISM. The wind predictions we have
incorporated here are based on the computations presented in \citet{arcones}.
Here we have tested the impact that wind parameters, such as entropy, electron
fraction, and expansion time scale, have on the neutron-capture nucleosynthesis
in the wind. The effect of progenitor mass and progenitor metallicity remains an
open question, since the supernova models still have uncertainties that are too large to
constrain these quantities. Nonetheless, we note that since this process is a
primary process, the impact of metallicity is not the most important factor when
trying to constrain the yields. Assuming a neutron-rich wind with the
electron fraction constricted to $0.4 < Y_e < 0.49$ \citep[as currently
suggested by][]{pinedo,Roberts12,roberts2}, we can try to loosely confine some parameters
to realistic ranges by setting an entropy interval of 50 to
150\,$k_B/\rm{baryon}$ and a wind expansion time scale limited to a few
milliseconds. These are typical values found in hydrodynamical wind simulations
\cite[]{arcones07,fischer10} and lead to Sr yields spanning $10^{-4} -
10^{-7}\,M_{\odot}$, where the largest contribution comes from 12 -- 25
$M_{\odot}$ supernovae, and the smallest yield could be assigned to 8 --
12 $M_{\odot}$ SN explosions. Generally speaking, the smaller Sr yields can be assigned to low-mass progenitors, while more massive ones produce larger Sr yields.

The first s-process yield we consider here are those of
\citet{bis2010}, who provide the yields from AGB stars in the mass range $1.3 - 2
M_{\odot}$ at different metallicities ([Fe/H]): 0, -0.8, -1.6, and -2.6 (S.
Bisterzo, priv. comm.). These yields have been calculated with the FRANEC (Frascati Phapson-Newton
Evolutionary Code) that uses reaction rates from the KADoNiS and NACRE
databases. The neutron-capture elements are created in $^{13}$C
pockets and brought to the stellar surface during thermal pulses. The AGB star
experiences a mass loss in the range $10^{-4} - 10^{-7} M_{\odot}/yr$ where
the interval between successive thermal pulses lasts for about $10^4 - 10^5$
years depending on the AGB core mass (e.g., see model data published in
\citealt{karakas} or \citealt{cristallo}).
The impact at low metallicity ([Fe/H]$<-2.6$), which we are interested in, is minor and would
need to be extrapolated. However, yields from low-mass AGB stars are important
at higher metallicities.

We also use AGB yields based on the nucleosynthesis calculations presented in
\citet{karakas} and \citet{lugaro}. These calculations use reaction rates from
the NACRE and JINA databases, which includes the KADoNIS neutron-capture cross
sections. In comparison to the yields from \citealt{bis2010}, these span a range
from 0.9$M_{\odot}$ to 6$M_{\odot}$ at [Fe/H] = -2.3; at higher metallicities we
cover a mass range from 1.25$M_{\odot}$ to 6$M_{\odot}$ at [Fe/H] = -0.15 and
5$M_{\odot}$ to 8$M_{\odot}$ at [Fe/H] = +0.14 (slightly super solar). For the
heaviest of these stars, the main neutron source is the
$^{22}$Ne($\alpha$,n)$^{25}$Mg reaction, and a $^{13}$C pocket has been introduced.
However, for the lowest mass stars, we introduce some partial mixing of protons
at the deepest extent of each third dredge-up episode in order to form a
$^{13}$C pocket \citep[see detailed discussion in][]{lugaro}. For most models
the mass-loss rate used on the AGB is the semi-empirically derived \citet{vassi}
mass loss formulae. One of the major uncertainties in AGB modelling is the
mass-loss rate, especially at the lowest metallicities, and for this reason we
experiment with variations to the mass loss in the intermediate-mass AGB models.
We refer to the discussions in \citet{karak2010} and \citet{karakas} for more
details.

The s-process yields of massive rapid rotating stars are from
\citet{frisch}; these results
are more recent than \citet{pignatari}, and \citet{frisch} based their yield
predictions on a stellar evolution code. The code used is the
  Geneva stellar evolution code (GENEC), with reaction libraries from
  REACLIB, and neutron captures from KADoNiS. Moreover, the results by \citet{frisch}
are based
on the still favoured reaction rate for $^{17}O(\alpha,\gamma)$ by
\cite{caughlan}. We explore the uncertainty
linked to this reaction rate decreasing it by a factor of 10. In this
way we can generate upper and lower limits for what we might expect as
Sr yields from these massive rotating stars. In general the
above-mentioned s-process yields are in the range
$10^{-6}-10^{-9}M_{\odot}$. (For comparison
  the non-rotating stars barely produce any Sr.)

\subsection{The impact of different stellar yields on Galactic chemical
evolution (GCE) models}

We now show a chemical evolution model computed with the
different assumptions for the stellar yields discussed above. Because the goal of
this section is to check for the broad impact of the different stellar yields on
the chemical enrichment, we adopt a homogeneous chemical evolution model to guide our discussion. For that we use the halo chemical evolution model of
\citet{chiappini08}, which we briefly summarise here:\\
 1) the infall law of the primordial gas follows a Gaussian function,\\
 2) the occurring outflow from the system is proportional to the star formation
rate (SFR).

The timescale for the formation of the halo is fast (less than 0.5 Gyr).
The other crucial element for our study is iron, for which we adopt the
predictions from \citealt[][(WW95)]{WW95}.  As shown in
\citet{cescut10}, this combination of parameters is able to produce
a synthetic metallicity distribution function, which is in good agreement
with the one observed in the Galactic halo \citep{lai,MDF}. As a result we
assume that this model follows the correct timescale for the
chemical enrichment.

We have not run individual models for each set of AGB nucleosynthesis
prediction. The reason is that, even though our chemical evolution
model calculates the enrichment from low-mass AGB stars and
SNe Ia, only the more massive stars (SN II) can be considered drivers of the
early
chemical enrichment in the halo. This is due to the considerable longer
timescales needed to evolve AGBs and SNe Ia. We have only tested the metal-poor
yields for Sr
\citep[i.e. from the metal-poor extension to the yields from][]{lugaro}, for
which we have
predictions up to 6$M_{\odot}$ stars (compared to the $2 - 3M_{\odot}$ from
\citealt{bis2010,cristallo}). The massive AGB stars are more likely to
contribute to the chemical enrichment during the formation of the halo, i.e. at lower metallicities.
Nevertheless, the enrichment produced by the 6$M_{\odot}$ AGB stars do not
influence the overall results. Only when extremely low yields from
even more massive stars (SN II) are adopted (lower than {\it Wlow}), is it possible to see the influence
of the heavy AGB yields. 

\begin{figure}
\centering
\vspace{0mm}
\includegraphics[width=0.43\textwidth]{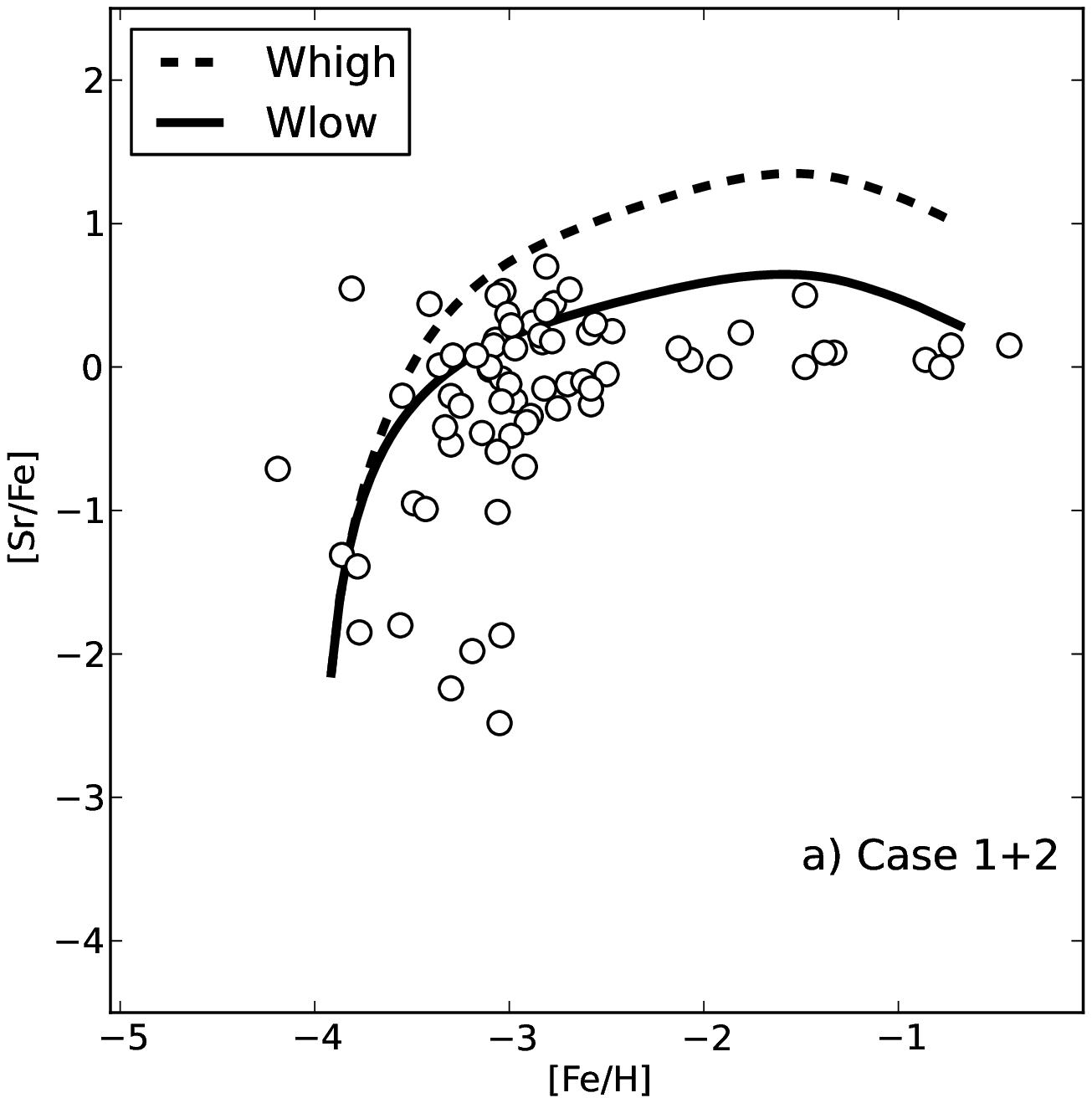}\\
\vspace{-3mm}
\includegraphics[width=0.43\textwidth]{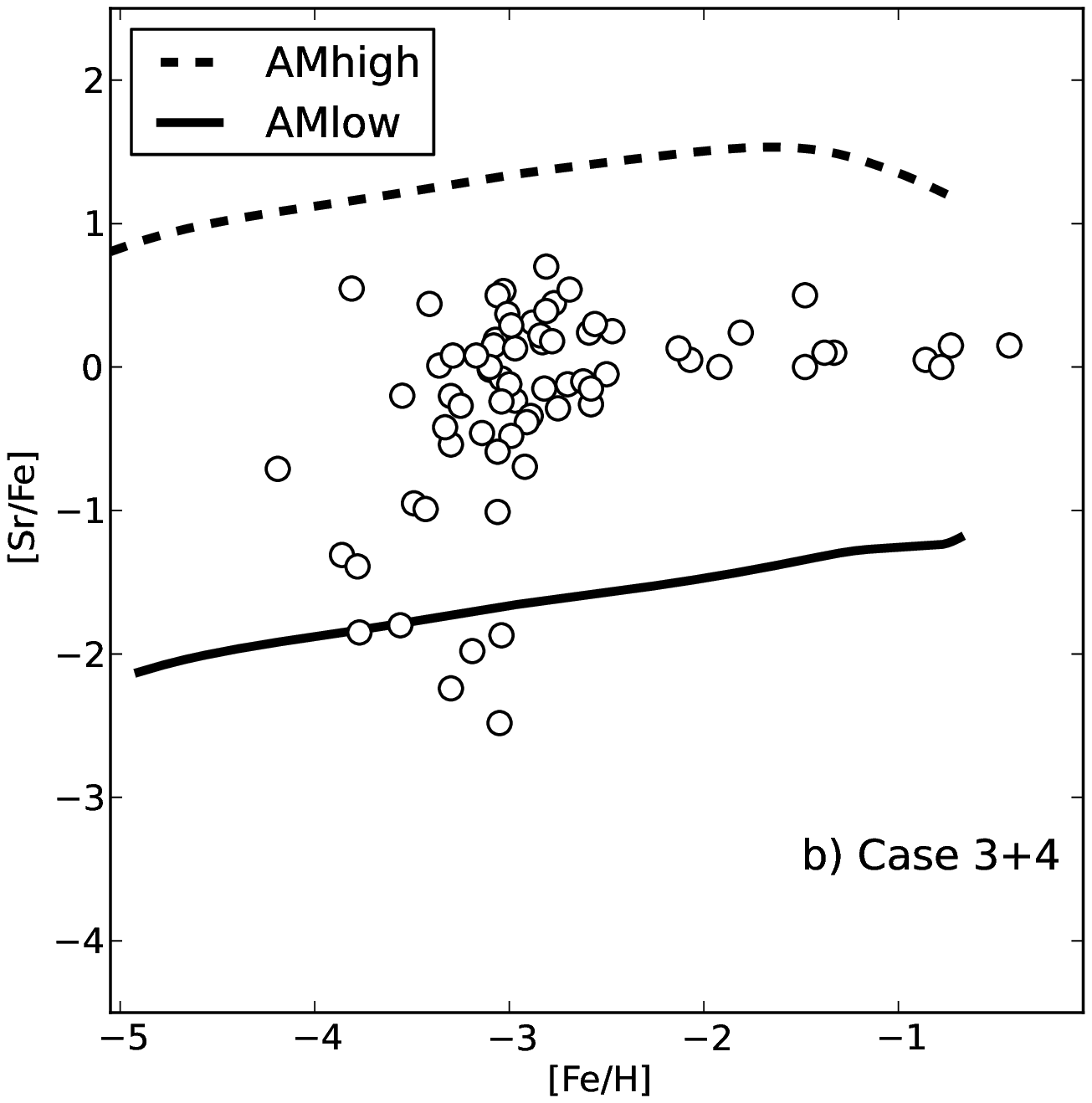}
\vspace{-4mm}
\includegraphics[width=0.43\textwidth]{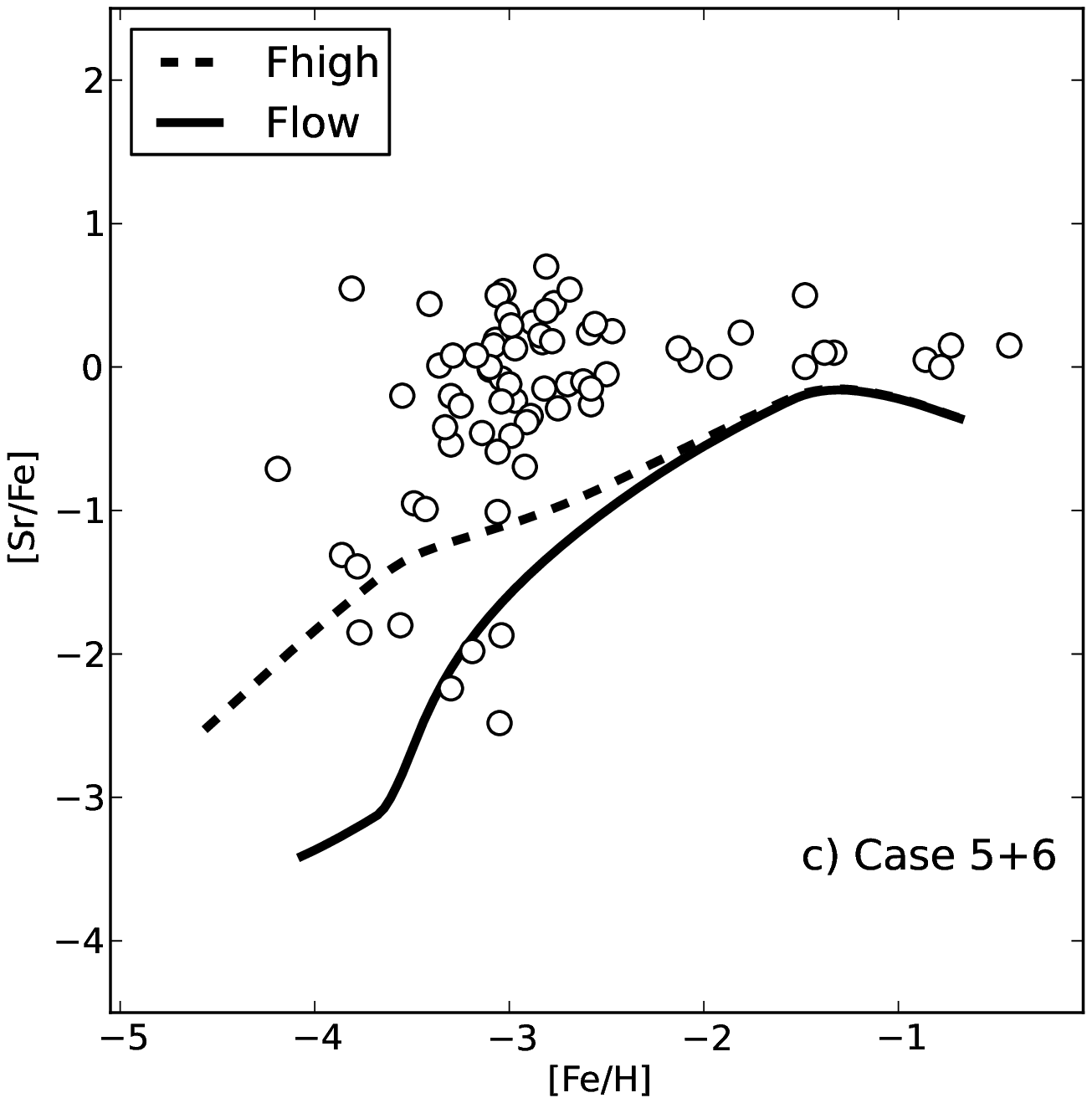}
\vspace{-2mm}
\caption{GCE model predictions compared to Case 1+2), which is based on ECSN yields from S.
Wanajo in the top panel a). The predictions are in the middle panel b) compared to Case 3+4) which is neutrino-driven winds from Arcones \& Montes, and in the bottom panel c) the comparison is made to Case 5+6), namely fast-rotating stars from U.
Frischknecht. The dashed/solid lines represent the upper/lower limits to the Sr yields,
respectively.}
\label{GCE}
\end{figure}

We have computed different chemical evolution models with different
nucleosynthesis assumptions, namely:
\begin{itemize}
\item Case 1) Wanajo low ({\it Wlow}): 1.79$\cdot10^{-4}M_{\odot}$ in the range
9.5-10$M_{\odot}$;\\
\item Case 2) Wanajo high ({\it Whigh}): 1.79$\cdot10^{-4}M_{\odot}$ in the range
8-10$M_{\odot}$;\\ 
\item Case 3) Arcones \& Montes low ({\it AMlow}): $10^{-7}M_{\odot}$ mass range
8-25$M_{\odot}$;\\
\item Case 4) Arcones \& Montes high ({\it AMhigh}): $10^{-4}M_{\odot}$ in mass range
8-25$M_{\odot}$\footnote{Arcones \& Montes standard ({\it Astand}): $10^{-4}M_{\odot}$
in mass range 12-25$M_{\odot}$ and $10^{-7}M_{\odot}$ in mass range
8-12$M_{\odot}$ and M$> 25M_{\odot}$.};\\
\item Case 5) Frischknecht low ({\it Flow}) with rotation and their 'standard' value for
the reaction rate, generalising the production of the 25$M_{\odot}$ in their
paper to a range of 15-40$M_{\odot}$; \\
\item Case 6) Frischknecht high ({\it Fhigh}) calculated as Flow but with a decreased value
for the reaction $^{17}$O rate (producing higher Sr yields)
\end{itemize}

In Fig. \ref{GCE}a) (top) we show the models {\it Wlow} and {\it Whigh}, which differ in the level of Sr enrichment by the ECSN due to the supernova mass used in the GCE model. The high level of Sr
production in a narrow range of masses produces in the case of Whigh an important bump in the [Sr/Fe]
predictions.
This roughly appears as a knee in the [Sr/Fe] trend at [Fe/H]$\sim -3$.
However, the {\it Whigh} model produces too much Sr compared to the
  observed values. Including CEMP-s stars, we would have found some stars
  closer to the {\it Whigh} model only at the lowest metallicity. 
  The {\it Wlow} model predicts [Sr/Fe] ratios very close to the observations at all [Fe/H]
    values.

Model {\it AMlow} and {\it AMhigh} are presented in Fig. \ref{GCE}b) (middle). In these models
the Sr is produced
in a wide range of massive stars, which leads to a rather flat [Sr/Fe] trend.
The nearly 3 dex difference between the
  upper and lower curves ({\it AMlow} and {\it AMhigh}) reflects the variations and uncertainties in the Sr yields.

From the results obtained from the two pairs of models we cannot draw strong
conclusions as to the nucleosynthetic origin of Sr in our metal-poor star
sample. The only message we can read from them is that some
  combination of two r-process sites, can roughly explain the observationally derived abundances.
At this level, the observations/data can still be used to constrain the
theoretical yields, rather than the other way around.

Finally, from the two models generated using the \citet{frisch} yields (Fig. \ref{GCE}c) bottom panel), 
we can conclude that the s-process from fast-rotating massive stars may be an important
source of Sr during the Galactic halo formation
\citep[see][]{ces13}. The early
s-process taking place in metal-poor, fast rotating, massive stars should be
coupled/added to at least one r-process.

\begin{figure}
\centering
\includegraphics[width=0.48\textwidth]{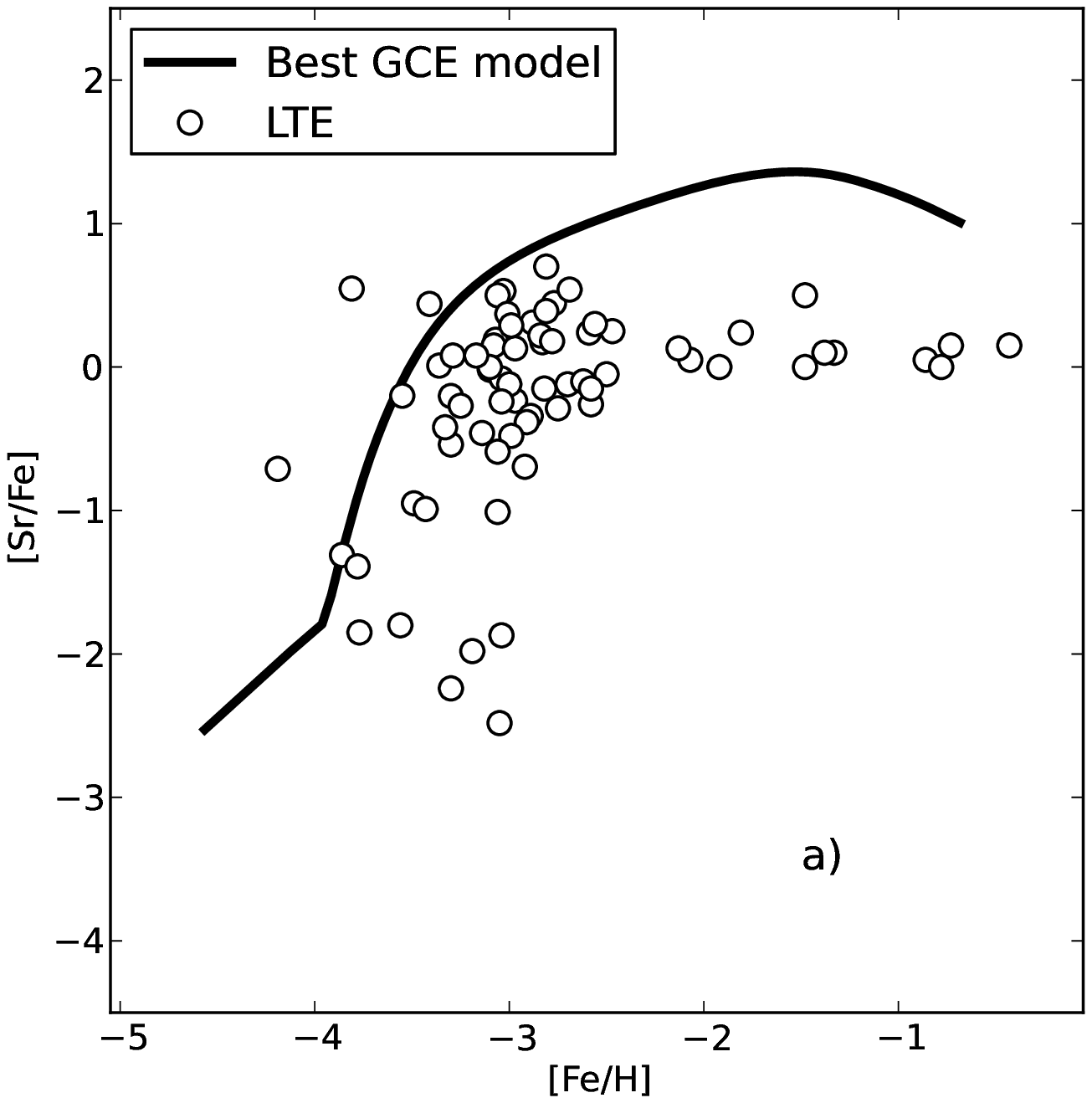}
\hspace{-5mm}
\includegraphics[width=0.48\textwidth]{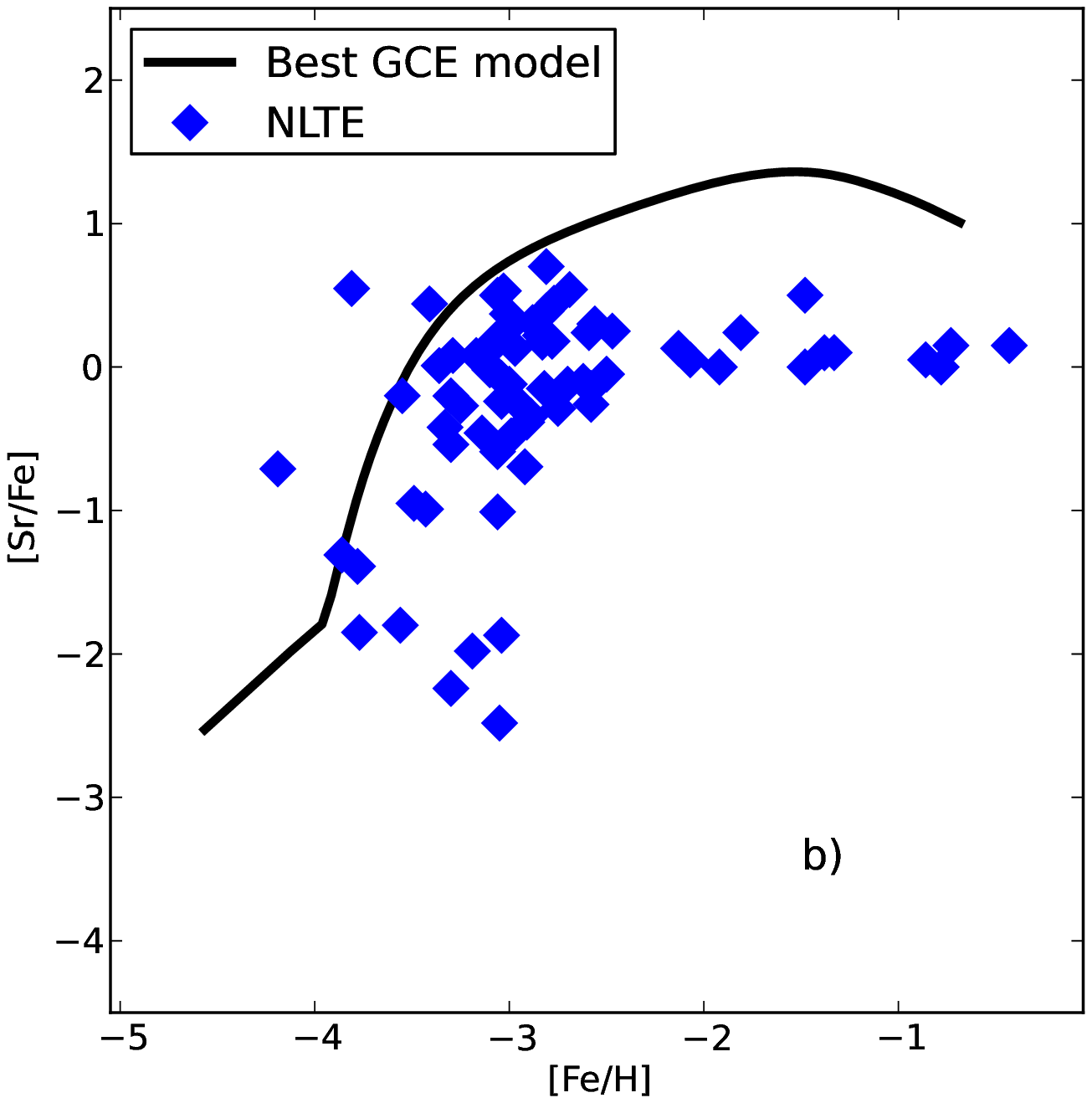}
\caption{Comparison of the best homogeneous GCE model (solid line) to Sr LTE abundances (open circles) in the upper panel a), and to NLTE Sr abundances (filled blue diamonds) in the lower panel b).}
\label{ltevsnlte}
\end{figure}

Since the difference between the LTE and NLTE Sr abundances is small, despite
the slight horizontal shift caused by the positive NLTE effect on
  metallicity, the methodology does not affect the
interpretations in a GCE context (see Fig. \ref{ltevsnlte}). Therefore only the same GCE conclusions can be
drawn under LTE and NLTE. We stress that this is only true for an element like
Sr. Owing to the combined NLTE effects
  in Sr and Fe, there are no stars below [Fe/H] =$-3.5$ with high
  strontium abundances ([Sr/Fe]$>0$). Our NLTE corrected
abundances still show a large star-to-star scatter as is also found in
\citet{andrSr}.
 
Without precise theoretical yields it is impossible to draw firm conclusions
based on the abundance measurements. On the other 
hand, the presence of the spread can be explained in a stochastic chemical
evolution scenario.

Many advances have been made in modelling AGB stars, SN, and the r-process, as
well as significant improvements in the determination of atomic data for heavy
elements. These have in turn improved the GCE models. However, the
r-process yields still need to be better constrained not to span 3\,dex. Until
these improvements are made, yield predictions should only be used as guiding
upper and lower limits. The yields from the fast-rotating stars and the AGB
stars also face challenges when trying to constrict the $^{13}$C pockets, pulse
duration, mass loss, and poisons in order to improve the networks and, in turn,
yields. Only within the last decade have SN models managed to explode, and the
treatment of $^{13}$C pockets and mass loss from AGB stars is improving
\citep[e.g.][]{mat,matt08,sloan}. 
For example, by using AGB stars in star clusters it is possible to
constrain uncertain parameters related to convection and mass loss
\citep[e.g.][]{lebzel,kamath}, while $^{13}$C pocket sizes
can be constrained using data from a variety of observational data
\citep[e.g.,][]{lug2003, BM07, bis12}.

There are still many challenges, including a detailed understanding of
the formation mechanism of $^{13}$C pockets, which is currently unknown, and
a significant uncertainty in s-process models of AGB stars \citep[e.g., see
discussion in][]{herwig}. The yields from fast-rotating massive
stars also face some of the same challenges including an accurate
description of mass loss and convection in stellar models \citep[e.g., see
review by][]{langer}.

However, on the basis of Figs. \ref{GCE} and \ref{ltevsnlte} we believe that at least two different
sources (e.g. neutrino-driven winds from massive stars or low-mass ECSN) are
needed to explain the [Sr/Fe] abundances we derived under both LTE and
NLTE. An early s-process could take place in fast-rotating
stars and provide a possible explanation for the low [Sr/Fe] we obtain
at very low metallicity.

\section{Conclusions}

Cross-disciplinary work is crucial for computing better yields and, in
turn, better chemical evolution models. Only then will such calculations allow us
to accurately predict the behaviour and evolution of heavy elements such as
Sr. This information is needed in the era of high-resolution surveys, where we have a large flow of data coming in. We are
now capable of analysing the stars and extracting very accurate abundances with
uncertainties $<0.25$ dex, compared to what can be obtained from nucleosynthetic
yields and evolved stellar models. This indicates that observationally
derived  abundances are more likely to constrain the parameter space
of the yield predictions and the GCE models than vice versa.

On the basis of this study we can conclude the following about Sr in metal-poor
stars.
\begin{itemize}

\item When deriving \ion{Sr}{I} abundances, NLTE corrections should always be
applied in order to obtain ionisation equilibrium.

\item In the interval $-3.0 <$ [Fe/H] $<\sim -1.0$ the chemical evolution of
  Sr, as derived from the \ion{Sr}{II} resonance lines, is similar
  under LTE and NLTE. However, below [Fe/H] $= -3.0$,
  the NLTE corrections to \ion{Sr}{ii} lines are important to
    obtain the correct Sr abundances.

\item The abundances obtained from the \ion{Sr}{ii} lines are
  sensitive to surface gravity. If the latter parameter is obtained
  from the ionisation equilibrium of Fe,  LTE approximation should not
  be used, because it leads to large systematic errors in $\log g$ of
  up to $+0.8$ dex. 

\item In their current state the Sr yields are too uncertain to clearly
disentangle contributions from different processes or sites. We may instead use
the observationally derived abundances to constrain the parameter space of the
model predictions.
\end{itemize}

In summary, it is not sufficient to account for NLTE effects in
  the line formation of the Sr lines. NLTE effects must be taken into
  account in determining stellar parameters, i.e., surface
  gravities and metallicities. Alternatively, abundances determined
  from the lines of the majority species, \ion{Sr}{ii} and
  \ion{Fe}{ii}, can be used. The LTE assumption is a trustworthy chemical evolution tracer, in the interval
   $-3 < $ [Fe/H] $ < 0$, for the Sr II abundances from dwarfs calculated with
   gravities stemming from parallaxes and temperatures based on accurate
   photometry.
   The metal-poor giants, which are the best targets at low
   metallicity, are biased by the LTE assumption. The 
   parameters of the giants and their abundances must be computed under NLTE. 

With the current uncertainties on the stellar yields, which span 
    two to three orders of magnitude, we cannot draw strong
    conclusions on the chemical evolution of Sr in the early
    Galaxy. Neither can we precisely extract the various sites that 
    contribute to the creation of the large star-to-star scatter, although at least two sites seem necessary. The yield
    predictions and the GCE model used in this work provide both upper and lower limits to the highly
    scattered stellar strontium abundances.

\begin{acknowledgements}

This work was supported by Sonderforschungsbereich SFB 881 "The Milky Way
System" (subproject A5) of the German Research Foundation (DFG). 
Based on observations made with the European Southern
Observatory telescopes
(obtained from the ESO/ST-ECF Science Archive Facility) and the Calar Alto
Observatory telescopes. We are grateful to R. Sch\"onrich for useful
discussions, to S. Bisterzo for yields, and to L. Sbordone, H.-T. Janka, and S. Wanajo for comments. We
thank T. Gehren for providing spectra, and S. Wanajo for ECSN yields. A.A.  is
supported by the Helmholtz-University Young
Investigator grant No. VH-NG-825. Finally, we thank the referee for the detailed comments.

\end{acknowledgements}
\bibliographystyle{aa}
\bibliography{Sr_nlte_RESUB}

\begin{thebibliography}{81}
\expandafter\ifx\csname natexlab\endcsname\relax\def\natexlab#1{#1}\fi

\bibitem[{{Alonso} {et~al.}(1996){Alonso}, {Arribas}, \&
  {Martinez-Roger}}]{alondw}
{Alonso}, A., {Arribas}, S., \& {Martinez-Roger}, C. 1996, \aap, 313, 873

\bibitem[{{Alonso} {et~al.}(1999){Alonso}, {Arribas}, \&
  {Mart{\'{\i}}nez-Roger}}]{AlonsoG}
{Alonso}, A., {Arribas}, S., \& {Mart{\'{\i}}nez-Roger}, C. 1999, \aaps, 140,
  261

\bibitem[{{Anders} \& {Grevesse}(1989)}]{anders}
{Anders}, E. \& {Grevesse}, N. 1989, \gca, 53, 197

\bibitem[{{Andrievsky} {et~al.}(2011){Andrievsky}, {Spite}, {Korotin}, {Fran{\c
  c}ois}, {Spite}, {Bonifacio}, {Cayrel}, \& {Hill}}]{andrSr}
{Andrievsky}, S.~M., {Spite}, F., {Korotin}, S.~A., {et~al.} 2011, \aap, 530,
  A105

\bibitem[{{Andrievsky} {et~al.}(2009){Andrievsky}, {Spite}, {Korotin}, {Spite},
  {Fran{\c c}ois}, {Bonifacio}, {Cayrel}, \& {Hill}}]{andrievsky09}
{Andrievsky}, S.~M., {Spite}, M., {Korotin}, S.~A., {et~al.} 2009, \aap, 494,
  1083

\bibitem[{{Arcones} {et~al.}(2007){Arcones}, {Janka}, \& {Scheck}}]{arcones07}
{Arcones}, A., {Janka}, H.-T., \& {Scheck}, L. 2007, aap, 467, 1227

\bibitem[{{Arcones} \& {Montes}(2011)}]{arcones}
{Arcones}, A. \& {Montes}, F. 2011, \apj, 731, 5

\bibitem[{{Beers} \& {Christlieb}(2005)}]{beersFe}
{Beers}, T.~C. \& {Christlieb}, N. 2005, \araa, 43, 531

\bibitem[{{Belyakova} \& {Mashonkina}(1997)}]{srnlte}
{Belyakova}, E.~V. \& {Mashonkina}, L.~I. 1997, Astronomy Reports, 41, 530

\bibitem[{{Bergemann} \& {Gehren}(2008)}]{berggeh}
{Bergemann}, M. \& {Gehren}, T. 2008, \aap, 492, 823

\bibitem[{{Bergemann} {et~al.}(2012{\natexlab{a}}){Bergemann}, {Hansen},
  {Bautista}, \& {Ruchti}}]{bergSr}
{Bergemann}, M., {Hansen}, C.~J., {Bautista}, M., \& {Ruchti}, G.
  2012{\natexlab{a}}, \aap, 546, A90

\bibitem[{{Bergemann} {et~al.}(2012{\natexlab{b}}){Bergemann}, {Lind},
  {Collet}, {Magic}, \& {Asplund}}]{MB1d3d}
{Bergemann}, M., {Lind}, K., {Collet}, R., {Magic}, Z., \& {Asplund}, M.
  2012{\natexlab{b}}, \mnras, 427, 27

\bibitem[{{Bisterzo} {et~al.}(2010){Bisterzo}, {Gallino}, {Straniero},
  {Cristallo}, \& {K{\"a}ppeler}}]{bis2010}
{Bisterzo}, S., {Gallino}, R., {Straniero}, O., {Cristallo}, S., \&
  {K{\"a}ppeler}, F. 2010, \mnras, 404, 1529

\bibitem[{{Bisterzo} {et~al.}(2012){Bisterzo}, {Gallino}, {Straniero},
  {Cristallo}, \& {K{\"a}ppeler}}]{bis12}
{Bisterzo}, S., {Gallino}, R., {Straniero}, O., {Cristallo}, S., \&
  {K{\"a}ppeler}, F. 2012, \mnras, 422, 849

\bibitem[{{Bona{\v c}i{\'c} Marinovi{\'c}} {et~al.}(2007){Bona{\v c}i{\'c}
  Marinovi{\'c}}, {Lugaro}, {Reyniers}, \& {van Winckel}}]{BM07}
{Bona{\v c}i{\'c} Marinovi{\'c}}, A., {Lugaro}, M., {Reyniers}, M., \& {van
  Winckel}, H. 2007, \aap, 472, L1

\bibitem[{{Bonifacio} {et~al.}(2000){Bonifacio}, {Monai}, \& {Beers}}]{bonEBV}
{Bonifacio}, P., {Monai}, S., \& {Beers}, T.~C. 2000, \aj, 120, 2065

\bibitem[{{Bonifacio} {et~al.}(2009){Bonifacio}, {Spite}, {Cayrel}, {Hill},
  {Spite}, {Fran{\c c}ois}, {Plez}, {Ludwig}, {Caffau}, {Molaro}, {Depagne},
  {Andersen}, {Barbuy}, {Beers}, {Nordstr{\"o}m}, \& {Primas}}]{bonifaciodw}
{Bonifacio}, P., {Spite}, M., {Cayrel}, R., {et~al.} 2009, \aap, 501, 519

\bibitem[{{Busso} {et~al.}(1999){Busso}, {Gallino}, \& {Wasserburg}}]{busso}
{Busso}, M., {Gallino}, R., \& {Wasserburg}, G.~J. 1999, \araa, 37, 239

\bibitem[{{Butler} \& {Giddings}(1985)}]{butler85}
{Butler}, K. \& {Giddings}, J. 1985, Newsletter on Analysis of Astronomical
  Spectra No. 9 (University College London)

\bibitem[{{Casagrande} {et~al.}(2010){Casagrande}, {Ram{\'{\i}}rez},
  {Mel{\'e}ndez}, {Bessell}, \& {Asplund}}]{casagra}
{Casagrande}, L., {Ram{\'{\i}}rez}, I., {Mel{\'e}ndez}, J., {Bessell}, M., \&
  {Asplund}, M. 2010, \aap, 512, A54

\bibitem[{{Caughlan} \& {Fowler}(1988)}]{caughlan}
{Caughlan}, G.~R. \& {Fowler}, W.~A. 1988, Atomic Data and Nuclear Data Tables,
  40, 283

\bibitem[{{Cescutti}(2008)}]{cescutti}
{Cescutti}, G. 2008, \aap, 481, 691

\bibitem[{{Cescutti} \& {Chiappini}(2010)}]{cescut10}
{Cescutti}, G. \& {Chiappini}, C. 2010, \aap, 515, A102

\bibitem[{{Cescutti} {et~al.}(2013){Cescutti}, {Chiappini}, {Hirschi},
  {Meynet}, \& {Frischknecht}}]{ces13}
{Cescutti}, G., {Chiappini}, C., {Hirschi}, R., {Meynet}, G., \&
  {Frischknecht}, U. 2013, \aap, Submitted,

\bibitem[{{Chiappini} {et~al.}(2008){Chiappini}, {Ekstr{\"o}m}, {Meynet},
  {Hirschi}, {Maeder}, \& {Charbonnel}}]{chiappini08}
{Chiappini}, C., {Ekstr{\"o}m}, S., {Meynet}, G., {et~al.} 2008, \aap, 479, L9

\bibitem[{{Chiappini} {et~al.}(2011){Chiappini}, {Frischknecht}, {Meynet},
  {Hirschi}, {Barbuy}, {Pignatari}, {Decressin}, \& {Maeder}}]{chiapSr}
{Chiappini}, C., {Frischknecht}, U., {Meynet}, G., {et~al.} 2011, \nat, 472,
  454

\bibitem[{{Collet} {et~al.}(2007){Collet}, {Asplund}, \& {Trampedach}}]{collet}
{Collet}, R., {Asplund}, M., \& {Trampedach}, R. 2007, \aap, 469, 687

\bibitem[{{Cristallo} {et~al.}(2011){Cristallo}, {Piersanti}, {Straniero},
  {Gallino}, {Dom{\'{\i}}nguez}, {Abia}, {Di Rico}, {Quintini}, \&
  {Bisterzo}}]{cristallo}
{Cristallo}, S., {Piersanti}, L., {Straniero}, O., {et~al.} 2011, \apjs, 197,
  17

\bibitem[{{Dekker} {et~al.}(2000){Dekker}, {D'Odorico}, {Kaufer}, {Delabre}, \&
  {Kotzlowski}}]{dekker}
{Dekker}, H., {D'Odorico}, S., {Kaufer}, A., {Delabre}, B., \& {Kotzlowski}, H.
  2000, in Proc. SPIE, Vol. 4008, 534

\bibitem[{{Dobrovolskas} {et~al.}(2012){Dobrovolskas}, {Kucinskas},
  {Andrievsky}, {Korotin}, {Mishenina}, {Bonifacio}, {Ludwig}, \&
  {Caffau}}]{Ba3d}
{Dobrovolskas}, V., {Kucinskas}, A., {Andrievsky}, S.~M., {et~al.} 2012, A\&A,
  540,

\bibitem[{{Fischer} {et~al.}(2010){Fischer}, {Whitehouse}, {Mezzacappa},
  {Thielemann}, \& {Liebend{\"o}rfer}}]{fischer10}
{Fischer}, T., {Whitehouse}, S.~C., {Mezzacappa}, A., {Thielemann}, F., \&
  {Liebend{\"o}rfer}, M. 2010, aap, 517, A80+

\bibitem[{{Fran{\c c}ois} {et~al.}(2007){Fran{\c c}ois}, {Depagne}, {Hill},
  {Spite}, {Plez}, {Beers}, {James}, {Barbuy}, {Cayrel}, {Andersen},
  {Bonifacio}, {Molaro}, {Nordstr{\"o}m}, \& {Primas}}]{francois}
{Fran{\c c}ois}, P., {Depagne}, E., {Hill}, V., {et~al.} 2007, \aap, 476, 935

\bibitem[{{Frischknecht} {et~al.}(2012){Frischknecht}, {Hirschi}, \&
  {Thielemann}}]{frisch}
{Frischknecht}, U., {Hirschi}, R., \& {Thielemann}, F.-K. 2012, \aap, 538, L2

\bibitem[{{Gehren} {et~al.}(2004){Gehren}, {Liang}, {Shi}, {Zhang}, \&
  {Zhao}}]{gehren04}
{Gehren}, T., {Liang}, Y.~C., {Shi}, J.~R., {Zhang}, H.~W., \& {Zhao}, G. 2004,
  \aap, 413, 1045

\bibitem[{{Gehren} {et~al.}(2006){Gehren}, {Shi}, {Zhang}, {Zhao}, \&
  {Korn}}]{gehren06}
{Gehren}, T., {Shi}, J.~R., {Zhang}, H.~W., {Zhao}, G., \& {Korn}, A.~J. 2006,
  \aap, 451, 1065

\bibitem[{{Gonz{\'a}lez Hern{\'a}ndez} {et~al.}(2010){Gonz{\'a}lez
  Hern{\'a}ndez}, {Bonifacio}, {Ludwig}, {Caffau}, {Behara}, \&
  {Freytag}}]{gonher}
{Gonz{\'a}lez Hern{\'a}ndez}, J.~I., {Bonifacio}, P., {Ludwig}, H.-G., {et~al.}
  2010, \aap, 519, A46

\bibitem[{{Grupp}(2004{\natexlab{a}})}]{gruppa}
{Grupp}, F. 2004{\natexlab{a}}, \aap, 420, 289

\bibitem[{{Grupp}(2004{\natexlab{b}})}]{gruppb}
{Grupp}, F. 2004{\natexlab{b}}, \aap, 426, 309

\bibitem[{{Gustafsson} {et~al.}(2008){Gustafsson}, {Edvardsson}, {Eriksson},
  {J{\o}rgensen}, {Nordlund}, \& {Plez}}]{Gus08}
{Gustafsson}, B., {Edvardsson}, B., {Eriksson}, K., {et~al.} 2008, \aap, 486,
  951

\bibitem[{{Hansen} {et~al.}(2012){Hansen}, {Primas}, {Hartman}, {Kratz},
  {Wanajo}, {Leibundgut}, {Farouqi}, {Hallmann}, {Christlieb}, \&
  {Nilsson}}]{hansen}
{Hansen}, C.~J., {Primas}, F., {Hartman}, H., {et~al.} 2012, \aap, 545, A31

\bibitem[{{Heil} {et~al.}(2009){Heil}, {Juseviciute}, {K{\"a}ppeler},
  {Gallino}, {Pignatari}, \& {Uberseder}}]{heil}
{Heil}, M., {Juseviciute}, A., {K{\"a}ppeler}, F., {et~al.} 2009, PASA, 26, 243

\bibitem[{{Herwig}(2005)}]{herwig}
{Herwig}, F. 2005, \araa, 43, 435

\bibitem[{{Hoffman} {et~al.}(1997){Hoffman}, {Woosley}, \& {Qian}}]{hoffman}
{Hoffman}, R.~D., {Woosley}, S.~E., \& {Qian}, Y.-Z. 1997, \apj, 482, 951

\bibitem[{{Kamath} {et~al.}(2012){Kamath}, {Karakas}, \& {Wood}}]{kamath}
{Kamath}, D., {Karakas}, A.~I., \& {Wood}, P.~R. 2012, \apj, 746, 20

\bibitem[{{K{\"a}ppeler} {et~al.}(1989){K{\"a}ppeler}, Beer, \&
  Wisshak}]{kaep89}
{K{\"a}ppeler}, F., Beer, H., \& Wisshak, K. 1989, RPPh, 52, 945

\bibitem[{{Karakas}(2010)}]{karak2010}
{Karakas}, A.~I. 2010, \mnras, 403, 1413

\bibitem[{{Karakas} {et~al.}(2012){Karakas}, {Garc{\'{\i}}a-Hern{\'a}ndez}, \&
  {Lugaro}}]{karakas}
{Karakas}, A.~I., {Garc{\'{\i}}a-Hern{\'a}ndez}, D.~A., \& {Lugaro}, M. 2012,
  \apj, 751, 8

\bibitem[{{Lai} {et~al.}(2008){Lai}, {Bolte}, {Johnson}, {Lucatello}, {Heger},
  \& {Woosley}}]{lai}
{Lai}, D.~K., {Bolte}, M., {Johnson}, J.~A., {et~al.} 2008, \apj, 681, 1524

\bibitem[{{Langer}(2012)}]{langer}
{Langer}, N. 2012, \araa, 50, 107

\bibitem[{{Lebzelter} {et~al.}(2008){Lebzelter}, {Lederer}, {Cristallo},
  {Hinkle}, {Straniero}, \& {Aringer}}]{lebzel}
{Lebzelter}, T., {Lederer}, M.~T., {Cristallo}, S., {et~al.} 2008, \aap, 486,
  511

\bibitem[{{Lind} {et~al.}(2012){Lind}, {Bergemann}, \& {Asplund}}]{lindfe}
{Lind}, K., {Bergemann}, M., \& {Asplund}, M. 2012, ArXiv e-prints

\bibitem[{{Lugaro} {et~al.}(2003){Lugaro}, {Davis}, {Gallino}, {Pellin},
  {Straniero}, \& {K{\"a}ppeler}}]{lug2003}
{Lugaro}, M., {Davis}, A.~M., {Gallino}, R., {et~al.} 2003, \apj, 593, 486

\bibitem[{{Lugaro} {et~al.}(2012){Lugaro}, {Karakas}, {Stancliffe}, \&
  {Rijs}}]{lugaro}
{Lugaro}, M., {Karakas}, A.~I., {Stancliffe}, R.~J., \& {Rijs}, C. 2012, \apj,
  747, 2

\bibitem[{{Mart{\'{\i}}nez-Pinedo} {et~al.}(2012){Mart{\'{\i}}nez-Pinedo},
  {Fischer}, {Lohs}, \& {Huther}}]{pinedo}
{Mart{\'{\i}}nez-Pinedo}, G., {Fischer}, T., {Lohs}, A., \& {Huther}, L. 2012,
  ArXiv e-prints

\bibitem[{{Masana} {et~al.}(2006){Masana}, {Jordi}, \& {Ribas}}]{masana}
{Masana}, E., {Jordi}, C., \& {Ribas}, I. 2006, \aap, 450, 735

\bibitem[{{Mashonkina} {et~al.}(1999){Mashonkina}, {Gehren}, \&
  {Bikmaev}}]{mash99}
{Mashonkina}, L., {Gehren}, T., \& {Bikmaev}, I. 1999, \aap, 343, 519

\bibitem[{{Matsuura} {et~al.}(2007){Matsuura}, {Zijlstra}, {Bernard-Salas},
  {Menzies}, {Sloan}, {Whitelock}, {Wood}, {Cioni}, {Feast}, {Lagadec}, {van
  Loon}, {Groenewegen}, \& {Harris}}]{mat}
{Matsuura}, M., {Zijlstra}, A.~A., {Bernard-Salas}, J., {et~al.} 2007, \mnras,
  382, 1889

\bibitem[{{Mattsson} {et~al.}(2008){Mattsson}, {Wahlin}, {H{\"o}fner}, \&
  {Eriksson}}]{matt08}
{Mattsson}, L., {Wahlin}, R., {H{\"o}fner}, S., \& {Eriksson}, K. 2008, \aap,
  484, L5

\bibitem[{{Nissen} {et~al.}(2007){Nissen}, {Akerman}, {Asplund}, {Fabbian},
  {Kerber}, {Kaufl}, \& {Pettini}}]{nissen07}
{Nissen}, P.~E., {Akerman}, C., {Asplund}, M., {et~al.} 2007, \aap, 469, 319

\bibitem[{{Nissen} {et~al.}(1997){Nissen}, {Hoeg}, \& {Schuster}}]{nissengrav}
{Nissen}, P.~E., {Hoeg}, E., \& {Schuster}, W.~J. 1997, in Hipparcos - Venice
  '97, Vol. 402 (ESA Special Publication), 225

\bibitem[{{Nissen} {et~al.}(2002){Nissen}, {Primas}, {Asplund}, \&
  {Lambert}}]{nissen02}
{Nissen}, P.~E., {Primas}, F., {Asplund}, M., \& {Lambert}, D.~L. 2002, \aap,
  390, 235

\bibitem[{{Nomoto}(1987)}]{nomoto87}
{Nomoto}, K. 1987, \apj, 322, 206

\bibitem[{{{\"O}nehag} {et~al.}(2009){{\"O}nehag}, {Gustafsson}, {Eriksson}, \&
  {Edvardsson}}]{Oenehag}
{{\"O}nehag}, A., {Gustafsson}, B., {Eriksson}, K., \& {Edvardsson}, B. 2009,
  \aap, 498, 527

\bibitem[{{Pignatari} {et~al.}(2010){Pignatari}, {Gallino}, {Heil}, {Wiescher},
  {K{\"a}ppeler}, {Herwig}, \& {Bisterzo}}]{pigna}
{Pignatari}, M., {Gallino}, R., {Heil}, M., {et~al.} 2010, \apj, 710, 1557

\bibitem[{{Pignatari} {et~al.}(2008){Pignatari}, {Gallino}, {Meynet},
  {Hirschi}, {Herwig}, \& {Wiescher}}]{pignatari}
{Pignatari}, M., {Gallino}, R., {Meynet}, G., {et~al.} 2008, \apjl, 687, L95

\bibitem[{{Ram{\'{\i}}rez} \& {Mel{\'e}ndez}(2005)}]{ramirez}
{Ram{\'{\i}}rez}, I. \& {Mel{\'e}ndez}, J. 2005, \apj, 626, 465

\bibitem[{{Reetz}(1999)}]{reetz}
{Reetz}, J. 1999, PhD thesis, LMU M\"unich

\bibitem[{{Roberts}(2012)}]{roberts2}
{Roberts}, L.~F. 2012, \apj, 755, 126

\bibitem[{{Roberts} \& {Reddy}(2012)}]{Roberts12}
{Roberts}, L.~F. \& {Reddy}, S. 2012, ArXiv e-prints

\bibitem[{{Sbordone} {et~al.}(2010){Sbordone}, {Bonifacio}, {Caffau}, {Ludwig},
  {Behara}, {Gonz{\'a}lez Hern{\'a}ndez}, {Steffen}, {Cayrel}, {Freytag},
  {van't Veer}, {Molaro}, {Plez}, {Sivarani}, {Spite}, {Spite}, {Beers},
  {Christlieb}, {Fran{\c c}ois}, \& {Hill}}]{sbordone}
{Sbordone}, L., {Bonifacio}, P., {Caffau}, E., {et~al.} 2010, \aap, 522, A26

\bibitem[{{Schlegel} {et~al.}(1998){Schlegel}, {Finkbeiner}, \&
  {Davis}}]{schlegel}
{Schlegel}, D.~J., {Finkbeiner}, D.~P., \& {Davis}, M. 1998, \apj, 500, 525

\bibitem[{{Sch{\"o}rck} {et~al.}(2009){Sch{\"o}rck}, {Christlieb}, {Cohen},
  {Beers}, {Shectman}, {Thompson}, {McWilliam}, {Bessell}, {Norris},
  {Mel{\'e}ndez}, {Ram{\'{\i}}rez}, {Haynes}, {Cass}, {Hartley}, {Russell},
  {Watson}, {Zickgraf}, {Behnke}, {Fechner}, {Fuhrmeister}, {Barklem},
  {Edvardsson}, {Frebel}, {Wisotzki}, \& {Reimers}}]{MDF}
{Sch{\"o}rck}, T., {Christlieb}, N., {Cohen}, J.~G., {et~al.} 2009, \aap, 507,
  817

\bibitem[{{Sloan} {et~al.}(2012){Sloan}, {Matsuura}, {Lagadec}, {van Loon},
  {Kraemer}, {McDonald}, {Groenewegen}, {Wood}, {Bernard-Salas}, \&
  {Zijlstra}}]{sloan}
{Sloan}, G.~C., {Matsuura}, M., {Lagadec}, E., {et~al.} 2012, \apj, 752, 140

\bibitem[{{Sneden} {et~al.}(2008){Sneden}, {Cowan}, \& {Gallino}}]{chrisrev}
{Sneden}, C., {Cowan}, J.~J., \& {Gallino}, R. 2008, \araa, 46, 241

\bibitem[{{Spite} {et~al.}(2012){Spite}, {Andrievsky}, {Spite}, {Caffau},
  {Korotin}, {Bonifacio}, {Ludwig}, {Fran{\c c}ois}, \& {Cayrel}}]{spiteCa}
{Spite}, M., {Andrievsky}, S.~M., {Spite}, F., {et~al.} 2012, \aap, 541, A143

\bibitem[{Taylor(1997)}]{taylor}
Taylor, J. 1997, An Introduction to Error Analysis (University Science Books)

\bibitem[{{Travaglio} {et~al.}(2004){Travaglio}, {Gallino}, {Arnone}, {Cowan},
  {Jordan}, \& {Sneden}}]{Trav}
{Travaglio}, C., {Gallino}, R., {Arnone}, E., {et~al.} 2004, \apj, 601, 864

\bibitem[{{Vassiliadis} \& {Wood}(1993)}]{vassi}
{Vassiliadis}, E. \& {Wood}, P.~R. 1993, \apj, 413, 641

\bibitem[{{Vogt} {et~al.}(1994){Vogt}, {Allen}, {Bigelow}, {Bresee}, {Brown},
  {Cantrall}, {Conrad}, {Couture}, {Delaney}, {Epps}, {Hilyard}, {Hilyard},
  {Horn}, {Jern}, {Kanto}, {Keane}, {Kibrick}, {Lewis}, {Osborne},
  {Pardeilhan}, {Pfister}, {Ricketts}, {Robinson}, {Stover}, {Tucker}, {Ward},
  \& {Wei}}]{vogt94}
{Vogt}, S.~S., {Allen}, S.~L., {Bigelow}, B.~C., {et~al.} 1994, in Proc. SPIE,
  Vol. 2198, 362

\bibitem[{{Wanajo} {et~al.}(2011){Wanajo}, {Janka}, \& {M{\"u}ller}}]{wanLet}
{Wanajo}, S., {Janka}, H.-T., \& {M{\"u}ller}, B. 2011, \apjl, 726, L15

\bibitem[{{Woosley} \& {Weaver}(1995)}]{WW95}
{Woosley}, S.~E. \& {Weaver}, T.~A. 1995, \apjs, 101, 181

\end{thebibliography}
\Online

\appendix
\section{Abundance Details}
Here we show the LTE and the NLTE abundances, as well as parameters. The
abundance is given for the lines 4077\AA\, and 4607\AA\, individually.
\begin{sidewaystable}[h!]
\begin{center}
\caption{Sr abundances for the sample of stars. }
\label{allabun}
\hspace{-9mm}
\begin{tabular}{l c c c c c c |c c| c c c c c c c}
\hline
\hline
LTE &  \multispan{3}{LTE parameters\hfill}& EW & \multispan{2}{$       $ Syn $        $} & \multispan{2}{$  $NLTE parameters\hfill}& NLTE EW& Syn & Corr.& Full NLTE& Syn  & Corr.& Full NLTE\\
Star       &   T     &   logg &  [Fe/H]& 4077 &   4077 &  4607  &     logg  &  [Fe/H]&   4077  & 4077&$\Delta$NLTE&4077& 4607& $\Delta$NLTE & 4607\\
\hline
HD134169   &   5930. &   3.98 & --      &   --  &   0.05 &  $-0.15$ &     3.98  & $-0.86$ &   --    & 0.05  & $-0.01$ &0.040  &    --  & -- & 0.09 \\ 
HD148816   &   5880. &   4.07 & --      &   --  &   0.00 &  $-0.25$ &     4.07  & $-0.78$ &   --    & 0.00  & $-0.01$ &$-0.01$&    --  & -- & 0.01 \\
HD184448   &   5765  &   4.16 & --      &   --  &   0.15 &  $-0.15$ &     4.16  & $-0.43$ &   --    & 0.15  & $-0.01$ &0.14   &    --  & -- & 0.05 \\         
HD3567     &   6035. &   4.08 & $-1.33$ & $-0.04$ &  0.1   & $<-0.3$&     4.08  & $-1.29$ &$-0.1$  &  0.14 &  $-0.02$& 0.12  &$<-0.35$ &   0.28 &$<-0.07$ \\
HD19445    &   5982. &   4.38 & $-2.13$ & $-0.05$ &  0.13  & $<0.14$&     4.38  & $-2.10$ &$-0.12$ &  0.11 &  $-0.05$& 0.06  &$<0.05*$&  0.34 & $<0.39*$ \\
HD106038   &   5950. &   4.33 & $-1.48$ &   0.65  &  0.5   & 0.33   &     4.33  & $-1.45$ &   0.6  &  0.5  &  $-0.02$& 0.48  &  0.3   &   0.3 &   0.6 \\
HD121004   &   5711. &   4.46 & $-0.73$ &   0.21  &  0.15  & 0.2    &     4.46  & $-0.71$ &   0.2  &  0.15 &  $-0.01$& 0.14  &  0.14  &   0.24&   0.38 \\
HD122196   &   6048. &   3.89 & $-1.81$ &   0.05  &  0.24  &  --    &     3.89  & $-1.75$ &$-0.01$ &  0.22 &  $-0.03$& 0.19  &   --   &   --  &     \\
HD122563   &   4665  &   1.65 & $-2.60$ & $-0.09$ & $-0.05$& $<-0.6$&     1.65  & $-2.50$ &$-0.09$ &$-0.05$&   0.0   &$-0.05$&$<-0.7 $&   0.45&   $<-0.25$ \\
HD140283   &   5777  &   3.70 & $-2.58$ & $-0.33$ & $-0.15$&  --    &     3.70  & $-2.38$ &$-0.54$ &$-0.36$&  $-0.01$&$-0.37$&  --    &   --  &   -- \\
HD175305   &   5100. &   2.70 & $-1.38$ & $-0.02$ &  0.1   & $-0.35$&     2.70  & $-1.34$ &$-0.06$ &  0.11 &   0.0   & 0.11  & $-0.4$ &  0.37 &  $-0.03$ \\
G6412      &   6464  &   4.30 & $-3.24$ & $-0.05$ &  0.0   & --     &     4.30  & $-3.12$ &   0.04 &  0.05 &   0.12  & 0.17  &  --    &   --  &   -- \\
G6437      &   6494. &   3.82 & $-3.17$ & $-0.09$ &  0.08  & --     &    $4.23$ & $-3.00$ &$-0.05$ &  0.08 &   0.09  & 0.17  &  --    &   --  &   -- \\
BD133442   &   6450. &   4.20 & $-2.56$ &  0.26   &  0.3   &  --    &     $4.42$& $-2.47$ &  0.14  &  0.27 &  $-0.06$& 0.21  &  --    &   --  &   -- \\
CS30312-059&   5021. &   1.90 & $-3.06$ &  0.46   &  0.5   &  --    &     $2.41$& $-2.89$ &  0.28  &  0.35 &  $-0.04$& 0.31  &  --    &   --  &   -- \\
CS31082-001&   4925. &   1.51 & $-2.81$ &  0.59   &  0.7   &  --    &     $2.05$& $-2.63$ &  0.65  &  0.62 &  $-0.02$& 0.6   &  --    &   --  &   -- \\
HD74462    &   4590. &   1.84 & $-1.48$ &  0.1    &  0.0   & $-0.5$ &     $1.98$& $-1.43$ &  0.04  &$-0.12$&  0.0    &$-0.12$& $-0.5$ & 0.34  &   $-0.16$ \\
HD126238   &   4900  &   1.80 & $-1.92$ &  0.02   &  0.0   &$-0.34$ &      2.02 & $-1.85$ &  0.03  &  0.05 &  0.0    & 0.05  & $-0.46$&  0.43 &   $-0.03$\\
HD126587   &   4950. &   1.90 & $-3.01$ &  0.39   &  0.37  & $<-0.05$&     $2.36$& $-2.86$ &  0.34  &  0.25 &  $-0.04$& 0.21  &$<-0.15$&  0.4  & $<0.25$  \\
HE0315+0000&   5050. &   2.05 & $-2.81$ &  0.17   &  0.39  & --     &     $2.47$& $-2.67$ &  0.03  &  0.26 &  $-0.03$& 0.23  & --     &   --  &   -- \\
HE1219-0312&   5100. &   2.05 & $-2.99$ &  0.22   &  0.29  & --     &     $2.58$& $-2.81$ &  0.08  &  0.17 &  $-0.05$& 0.12  &  --    &   --  &   -- \\
\hline \hline
\end{tabular}
\tablefoot{
\tablefoottext{*}{Larger uncertainty $\pm0.1$\,dex in measurement}\\
\tablefoottext{<}{upper limit on the abundance}\\ 
}
\end{center}
\end{sidewaystable}
\clearpage

\end{document}